\documentclass[useAMS,usenatbib]{mn2e}

\usepackage{times}
\usepackage{amsfonts,amssymb}
\usepackage{aas_macros}
\usepackage[pdftex]{graphicx}
\usepackage[usenames,dvipsnames,svgnames,hyperref]{xcolor}

\title[nIFTy VI: dynamical substructure \& gaseous cluster outskirts]
      {nIFTy galaxy cluster simulations VI: the dynamical imprint of
        substructure on gaseous cluster outskirts.}
\author[Power et al.]
       {\parbox{\textwidth}{C. Power$^{1,2,3}$\thanks{E-mail: \texttt{chris.power@icrar.org}},
           P. J. Elahi$^{1,2}$, C. Welker$^{1,2}$, A. Knebe$^{4,5,1}$, F. R. Pearce$^6$,
           G. Yepes$^{4,5}$, R. Dav\'{e}$^{7}$, S. T. Kay$^{8}$, I. G. McCarthy$^{9}$,
           E. Puchwein$^{10}$, S. Borgani$^{11}$, D. Cunnama$^{12,13}$,
           W. Cui$^{4,1,3}$, \& J. Schaye$^{14}$}\vspace{0.4cm}\\
         $^1$ ICRAR, University of Western Australia, 35 Stirling Highway,
         Crawley, Western Australia 6009, Australia\\
         $^2$ ARC Centre of Excellence for All-Sky Astrophysics in 3 Dimensions (ASTRO 3D)\\
         $^3$ ARC Centre of Excellence for All-Sky Astrophysics (CAASTRO)\\
         $^4$ Departamento de F\'isica Te\'orica, M\'odulo 8, Facultad de Ciencias,
         Universidad Aut\'onoma de Madrid, 28049 Madrid, Spain\\
         $^5$ Centro de Investigaci\'{o}n Avanzada en F\'isica Fundamental (CIAFF),
         Facultad de Ciencias, Universidad Aut\'{o}noma de Madrid, 28049 Madrid, Spain\\         
         $^6$ School of Physics \& Astronomy, University of Nottingham, Nottingham
         NG7 2RD, UK\\
         $^7$ Institute for Astronomy, Royal Observatory, Edinburgh EH9 3HJ, UK\\         
         $^8$ Jodrell Bank Centre for Astrophysics, School of Physics and Astronomy,
         The University of Manchester, Manchester M13 9PL, UK \\
         $^{9}$ Astrophysics Research Institute, Liverpool John Moores University,
         146 Brownlow Hill, Liverpool L3 5RF, UK\\
         $^{10}$ Leibniz-Institut f\"{u}r Astrophysik Potsdam (AIP), An der Sternwarte 16, D-14482 Potsdam, Germany\\
         $^{11}$ INAF - Osservatorio Astronomico di Trieste, via G.B. Tiepolo 11,
         I-34143 Trieste, Italy\\
         $^{12}$ South African Astronomical Observatory, PO Box 9, Observatory, Cape Town 7935,
         South Africa\\
         $^{13}$ Department of Physics and Astronomy, University of the Western Cape, Cape Town 7535,
         South Africa\\
         $^{14}$ Leiden Observatory, Leiden University, PO Box 9513, NL-2300 RA Leiden,
         The Netherlands}
\begin{document}

\date{}

\pagerange{\pageref{firstpage}--\pageref{lastpage}} \pubyear{}

\maketitle

\label{firstpage}

\begin{abstract}
  Galaxy cluster outskirts mark the transition region from the mildly
  non-linear cosmic web to the highly non-linear, virialised, cluster interior.
  It is in this transition region that the intra-cluster medium (ICM) begins
  to influence the properties of accreting galaxies and groups, as ram pressure
  impacts a galaxy's cold gas content and subsequent star formation rate.
  Conversely, the thermodynamical properties of the ICM in this transition
  region should also feel the influence of accreting substructure (i.e.
  galaxies and groups), whose passage can drive shocks. In this paper, we use
  a suite of cosmological hydrodynamical zoom simulations of a single galaxy
  cluster, drawn from the {\small nIFTy} comparison project, to study how
  the dynamics of substructure accreted from the cosmic web influences the
  thermodynamical properties of the ICM in the cluster's outskirts. We
  demonstrate how features evident in radial profiles of the ICM (e.g. gas
  density and temperature) can be linked to strong shocks, transient and
  short-lived in nature, driven by the passage of substructure.
  The range of astrophysical codes and galaxy formation models in our
  comparison are broadly consistent in their predictions (e.g. agreeing
  when and where shocks occur, but differing in how strong shocks will be);
  this is as we would expect of a process driven by large-scale gravitational
  dynamics and strong, inefficently radiating, shocks. This suggests that
  mapping such shock structures in the ICM in a cluster's outskirts (via e.g.
  radio synchrotron emission) could provide a complementary measure of its recent
  merger and accretion history.
  
\end{abstract}

\begin{keywords}
  methods: numerical -- galaxies: clusters: general -- galaxies: formation
  -- galaxies: evolution -- cosmology: theory
\end{keywords}

\section{Introduction}
\label{sec:intro}
Galaxy clusters are the most massive virialised objects in the Universe, and
are widely used as a powerful testbed for our theories of dark matter, dark
energy, and galaxy formation and evolution, as well as cosmological parameter
estimation \citep[e.g.,][]{2012ARA&A..50..353K,2014MNRAS.440.2077M,
  2016MNRAS.456.4020M,
  2016MNRAS.459.1764S}.
Furthermore, they are striking signposts of the cosmic web, anchoring the
large-scale network of filaments, sheets, and voids
\citep[e.g.,][]{2011MNRAS.418..960V,2016A&A...588A..69D,2019MNRAS.486.3766M}.

The imprint of this cosmic web can be inferred from the spatial and kinematical
distribution of cluster galaxies
\citep[e.g.,][]{2004ApJ...603....7K,2005ApJ...627..647B,2005MNRAS.358..256P,2011ApJ...740...39H,2012ApJ...748...98S,2015MNRAS.450.2727T}.
However, its influence should also be evident in the presence of
accretion-driven shocks in the intra-cluster medium (ICM), especially at large
cluster-centric radius
\citep[e.g.,][]{2000ApJ...542..106T,2003ApJ...593..599R,2011MNRAS.418..960V}.
Such shocks arise naturally from cosmological accretion of gas onto the cluster
from the warm ($T\lesssim 10^4-10^6\,{\rm K}$) diffuse inter-galactic medium
(IGM) that resides in its lower density environs (see, e.g., 
\citealt{2001ApJ...552..473D}, and more recently \citealt{2018MNRAS.473...68C}
and \citealt{2019MNRAS.486.3766M}). This gas
will be accelerated gravitationally to peculiar velocities of order
$\sim\!1000\,{\rm km}\,{\rm s}^{-1}$ during infall, which will produce strong
shocks \citep[e.g.,][]{2018MNRAS.476...56Z} that give rise to various forms of
non-thermal emission, ranging from high energy X-rays and $\gamma$-rays
\citep[e.g.,][]{2003ApJ...585..128K,2004ApJ...617..281K,2014MNRAS.438..116Z},
to radio continuum emission in the form of synchrotron radiation as electrons
spiral along magnetic fields \citep{2008MNRAS.391.1511H,2009MNRAS.395.1333V}.

Both observations and simulations show trends between a cluster's merging
and mass accretion history - which, observationally, are inferred from
estimates of its dynamical state -  and properties of its ICM, such as gas
ellipticity \citep[e.g.,][]{2019arXiv190308662C}, turbulent
motions
\citep[e.g.,][]{2013ApJ...777..137N,2016ApJ...833..227A,2017ApJ...849...54L,2018MNRAS.481.1075S},
and the presence of density and pressure inhomogeneities
\citep[e.g.,][]{2011ApJ...731L..10N,2013MNRAS.428.3274Z,2013MNRAS.432.3030R,2015ApJ...806...68L,2016A&A...595A..42T}. This is of particular interest because these trends
contribute to biases in cluster mass estimates \citep[e.g.][]{2013MNRAS.428.3274Z};
for example, gas turbulence and accelerations undermine assumptions of
hydro-static equilibrium, which can underestimate a cluster's true mass
by between $\sim 5$ and 20 per cent \citep[e.g.,][]{2014ApJ...782..107N}, while
inhomogeneities drive clumping factors that boost estimates of the
X-ray emissivity \citep[e.g.][]{2017MNRAS.467.3827P,2018A&A...614A...7G},
which can be up to 40 per cent in the cluster outskirts, leading to
cluster masses being underestimated by up to $\sim 30$ per cent
\citep[e.g.,][]{2014ApJ...791...96R}.

However, examining how substructure impacts properties of the ICM is of
more general, astrophysical, interest. Examining the nature and origin
of these trends between a cluster's assembly history, its dynamical state,
and evidence of inhomogeneities in its ICM, can help to clarify our
understanding of how different physical processes can shape the ICM. For
example, what are the relative contributions to turbulence of, for example,
AGN jets and outflows compared to infalling substructures, and how does this
change with position within the cluster and its dynamical age? Developing
understanding in this way can provide theoretical insights into potentially new
and powerful observable signatures and tests of cluster assembly.

\medskip

Previous theoretical studies have investigated how merging and accretion
influence the thermodynamical structure of the ICM
\citep[e.g.,][]{2016ApJ...833..227A} using a statistical approach to
infer the likely contribution of substructure
\citep[such as, e.g., using the density distribution in radial bins to decompose
  the ICM into smooth and clumpy components; cf.][]{2013MNRAS.428.3274Z}, and
comparing to the degree to which a cluster is considered dynamically relaxed
\citep[e.g.][]{2013MNRAS.428.3274Z,2014ApJ...791...96R,2018PASJ...70...51O,2018MNRAS.476...56Z}
or by quantifying its assembly history by measuring mass growth over time
\citep[e.g.][]{2014ApJ...782..107N,2015ApJ...806...68L}. In this study,
we take a complementary approach; we look in detail at how the passage of 
substructures through the ICM, tracked from infall from the cosmic web,
influence its thermodynamical properties, as inferred from radial profiles
of density, temperature, and radial velocity. In particular, we focus our
analysis on ICM properties in the cluster outskirts, between $0.3\,R_{200}$
and $\sim(2-3)\,R_{200}$. It is here that we expect the signature of
mergers and accretion to be most pronounced
\citep[see, e.g.,][]{2015ApJ...806...68L} - we reason that infalling
substructures are likely to drive shocks into their surroundings when
they are at their most gas-rich, subsequent to their accretion onto the
cluster from the cosmic web - and we expect the effects of feedback
to be secondary.

We base our analysis on a suite of cosmological zoom simulations of a single
massive galaxy cluster, run with a range of state-of-the-art astrophysical
codes - spanning mesh- and particle-based hydrodynamics solvers, as well as
galaxy formation prescriptions - as part of the {\small nIFTy} galaxy cluster
comparison project \citep[see][and subsequent papers]{2016MNRAS.457.4063S}.
Previous papers in this series
\citep[e.g.,][]{2016MNRAS.457.4063S,2016MNRAS.458.4052C,2016MNRAS.tmp..582S}
have demonstrated that there are significant differences between code
predictions in the inner parts of clusters, reflecting both the choice of
hydrodynamics solver and the adopted prescription for galaxy formation. This
allows us to assess if ICM properties in the outskirts are similarly affected.
If shocks driven by infalling substructures are driving turbulence and
inhomogeneities in density and pressure, then we might expect systematic
differences in the strength and duration of shocks because of the choice of
hydrodynamics solver - as has been suggested by the work of
\citet{2014ApJ...791...96R} - but broad agreement in the timing and locations
of shocks because substructure orbits will be governed by the large-scale
gravitational field.

\medskip

The structure of this paper is as follows. In \S\ref{sec:simulated_cluster}, we
describe briefly the simulations we have used, including how they were set up and
their bulk properties at $z$=0. In \S\ref{sec:results}, we present our main
results -- connecting features evident in the radial profiles of ICM properties
(gas density, temperature, and kinematics) with substructure in
\S~\ref{ssec:spherical_averages}; exploring how the dynamics of substructure is
imprinted on the ICM in \S~\ref{ssec:mergers}; and demonstrating
the influence of the cosmic web in \S~\ref{ssec:lss}. Finally, we summarise
our results in \S\ref{sec:summary}.

\section{The Data}
\label{sec:simulated_cluster}

\subsection{The simulations}
Our analysis focuses on cosmological hydrodynamical zoom simulations of
a single galaxy cluster -- ``Cluster 19'' from the {\small{MUSIC}}
suite\footnote{\texttt{http://music.ft.uam.es/}}
\citep[][]{2013MNRAS.429..323S,2014MNRAS.439..588B,2014MNRAS.440.3520S} and
identified originally in the {\small{MultiDark}}\footnote{\texttt{www.cosmosim.org}}
$2048^3$ particle parent cosmological $N$-body simulation \citep[][]{2012MNRAS.423.3018P}.
The adopted cosmology assumes total matter, baryon, and dark energy density
parameters $(\Omega_{\rm m},\Omega_{\rm b},\Omega_{\Lambda})$=(0.27,0.0469,0.73); a
power spectrum normalisation of $\sigma_8$=0.82; a primordial spectral index
of $n$=0.95; and a dimensionless Hubble parameter of $h$=0.7, all in accord with the
WMAP7+BAO+SNI dataset of \citet{2011ApJS..192...18K}.

Initial conditions for all of the {\small{MUSIC}} clusters were generated
using the approach of \cite{2001ApJ...554..903K}, which can be summarised as follows;

\begin{enumerate}
\item All particles within a sphere of radius $6 h^{-1} {\rm Mpc}$ at
  $z$=0 centred on the cluster in the parent {\small{MultiDark}} simulation are found in
  the low-resolution $256^3$ particle version of the parent.
  
\item These particles are mapped back to the parent's initial conditions to identify
  the Lagrangian region from which they originated, from which a mask is created.

\item The initial conditions of the parent simulation are re-generated on a finer mesh
  of size $4096^3$, a factor of 8 increase in mass resolution relative to the parent
  simulation.
  
\item The mask is then applied to the $4096^3$ particle dataset, such that particles
  within the masked region are retained and those outside of the mask are binned to
  produce coarser mass resolution tidal particles, equivalent to the low-resolution
  $256^3$ version of the parent.
\end{enumerate}

\noindent The result is a set of initial conditions in which particles in the high
resolution patch have a mass resolution a factor of 8 higher than in the parent run,
corresponding to dark matter and gas particle masses of
m$_{\rm DM} = 9.01 \times 10^8 h^{-1} {\rm M}_{\odot}$ and
m$_{\rm gas} = 1.9 \times 10^8 h^{-1} {\rm M}_{\odot}$ respectively.

We used a selection of astrophysical codes - {\small Arepo}, {\small G2-X},
{\small G3-MUSIC}, {\small G3-OWLS}, and {\small G3-X} - and their associated
galaxy formation models (non-radiative and full physics) in this paper; in the
cases of {\small Arepo} and {\small G3-MUSIC}, we included two variants of
their galaxy formation models, with ({\small Arepo-IL}) and without AGN
feedback - ({\small Arepo-SH}, {\small G3-MUSIC-SH}, and {\small G3-MUSIC-PS}).
Key features are presented in the Appendix, and we refer the interested reader
to \citet{2016MNRAS.tmp..582S} for more details. We use only codes that have
used the same aligned parameters
\citep[see the Table 4 in][]{2016MNRAS.457.4063S} to re-simulate the selected
cluster; these govern the accuracy of the gravity solvers, and by requiring
alignment we ensure that cluster features that are driven by gravity - for
example, large-scale filamentary structure, orientation, timing of merger and
accretion events - should be consistent between runs. This allows us to focus
differences between results that are driven by either the hydrodynamics solver
or the adopted galaxy formation prescription.

\subsection{Structure finding}
We have used the phase-space structure-finder {\sc VELOCIraptor}\footnote{{\sc VELOCIraptor}
  derives from {\sc STructure Finder} \citep[see][]{2011MNRAS.418..320E} and can be
  downloaded from \texttt{https://github.com/pelahi/VELOCIraptor-STF.git}.}
\citep{2019arXiv190201010E} to identify the main cluster and its substructure.
{\sc VELOCIraptor} identifies particle groups using a 3 dimensional friends-of-friends
(FOF) algorithm and then refines each 3D-FOF group using a FOF algorithm applied to
the full 6 dimensional phase-space. This 6D-FOF group catalogue is cleaned to correct
for sets of halos that are combined artificially into a more massive system, arising
from spurious bridges of particles between halos that occur during the early stages of
mergers; this is done using the velocity dispersions of the 3D-FOF groups. Each 6D-FOF
halo is then decomposed into a smooth background and its substructures by applying the
phase-space FOF algorithm recursively, locating sets of particles that are dynamically
distinct from the background. This approach is capable of finding both subhaloes and
streams associated with tidally disrupted systems \citep[][]{2013MNRAS.433.1537E}.

\begin{table}
  \begin{center}
    \caption{\textbf{Cluster Properties at $z$=0.} Here we give
      (1) the virial mass, ${M}_{200}$, as defined in the text;
      (2) $R_{200}$, the corresponding virial radius;
      (3) $f_{\rm g}$, the fraction of ${\rm M}_{200}$ in gas;
      (4) the centre-of-mass offset $\Delta R=(\vec{R}_{\rm cen}-\vec{R}_{\rm cm})/R_{200}$; and
      (5,6) $\sigma_{\rm 3D}^{\rm d}$, and $\sigma_{\rm 3D}^{\rm g}$, the 3D velocity dispersions
      of all the dark matter and gas particles within ${\rm R}_{200}$.}
\vspace*{0.3 cm}

\begin{tabular}{cccccc}\hline
   & $M_{200}$   & $R_{200}$ & $f_{\rm g}$& $\Delta R$ & $\sigma_{\rm 3D}^{\rm d,g}$ \\
   & [$h^{-1} {\rm M}_{\odot}$] & [$h^{-1} {\rm Mpc}$]& &&[km/s]\\
  \hline
   & $1.107 \times 10^{15}$ &  1.682 & 0.1629 & 0.04 & 1960 / 950 \\  
\hline
\end{tabular}
\label{tab:cluster_props}
\end{center}
\end{table}

Cluster 19 has been studied extensively as part of the {\small nIFTy} galaxy cluster
comparison project \citep[cf.][and subsequent papers]{2016MNRAS.457.4063S}, and its
$z$=0 properties in the fiducial {\small G3-MUSIC} Full Physics run are given in
Table~\ref{tab:cluster_props}.
We follow the standard convention of defining cluster mass as $M_{200}$, the mass
enclosed within a spherical region of radius $R_{200}$ encompassing an
overdensity of 200 times the critical density $\rho_{\rm crit}$ at the
given redshift $z$, i.e.
\begin{equation}
  M_{200}=\frac{4\pi}{3} 200 \rho_{\rm crit}R_{200}^3.
\end{equation}
Following previous studies \citep[e.g.,][]{1998MNRAS.296.1061T,2012MNRAS.419.1576P},
we use the centre-of-mass offset to quantify the cluster's dynamical state,
\begin{equation}
  \Delta R=(\vec{R}_{\rm cen}-\vec{R}_{\rm cm})/R_{200},
\end{equation}
where $\vec{R}_{\rm cm}$ is the centre of mass of material within $R_{200}$
and $\vec{R}_{\rm cen}$ corresponds approximately to the centre of the densest
substructure within $R_{200}$ as calculated via the iterative method of
\citet{2003MNRAS.338...14P}. The measured value of $\Delta R$=0.04 indicates
that the cluster is relatively dynamically relaxed, which is consistent with
a visual impression of the cluster (see below).

\subsection{Visual Impression}
\label{ssec:visual_impression}

\begin{figure*}
  \centerline{
    \includegraphics[width=0.49\textwidth]{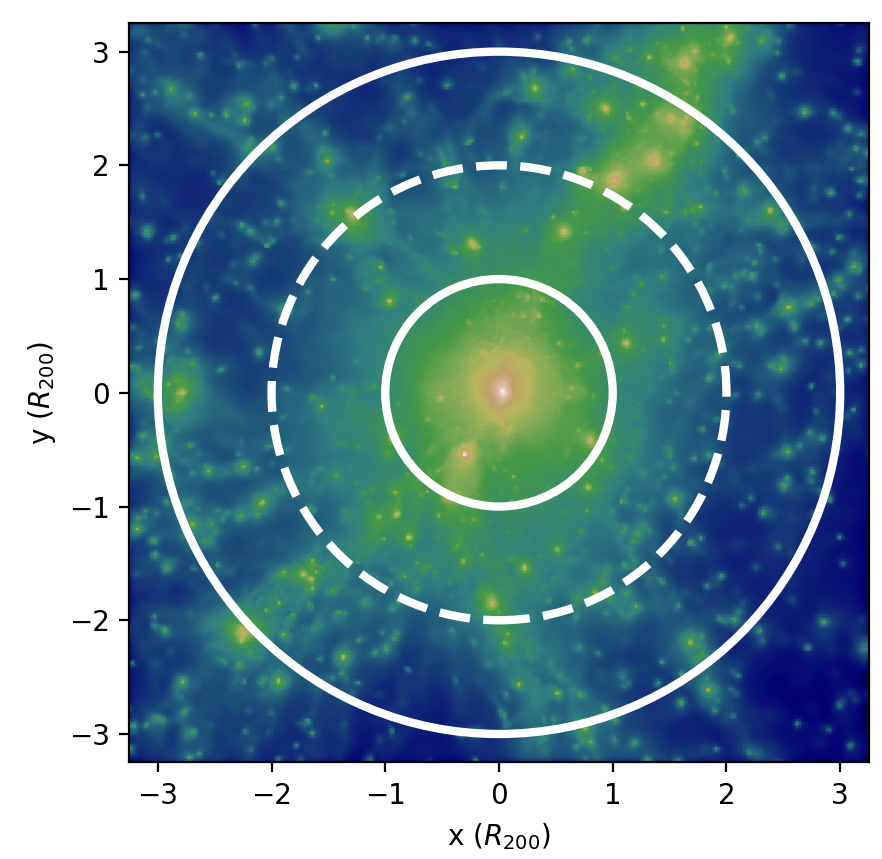}
    \includegraphics[width=0.48\textwidth]{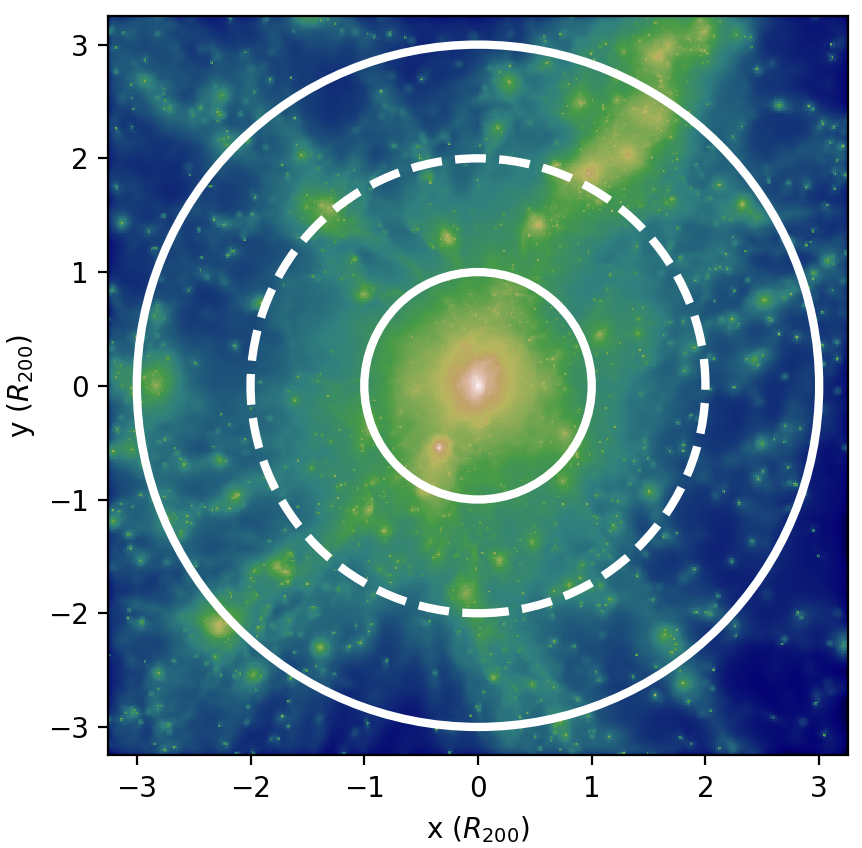}    
  }
  \centerline{
    \includegraphics[width=0.49\textwidth]{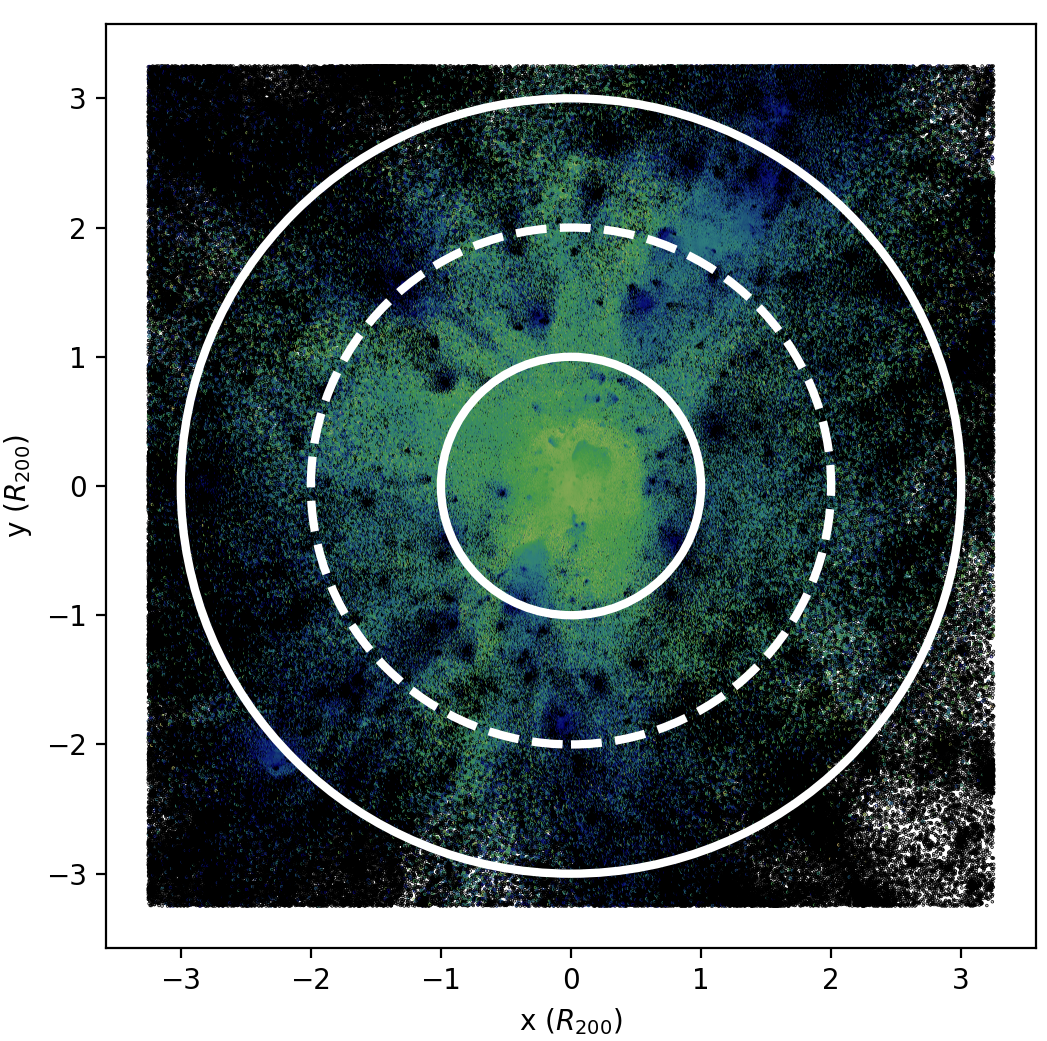}
    \includegraphics[width=0.49\textwidth]{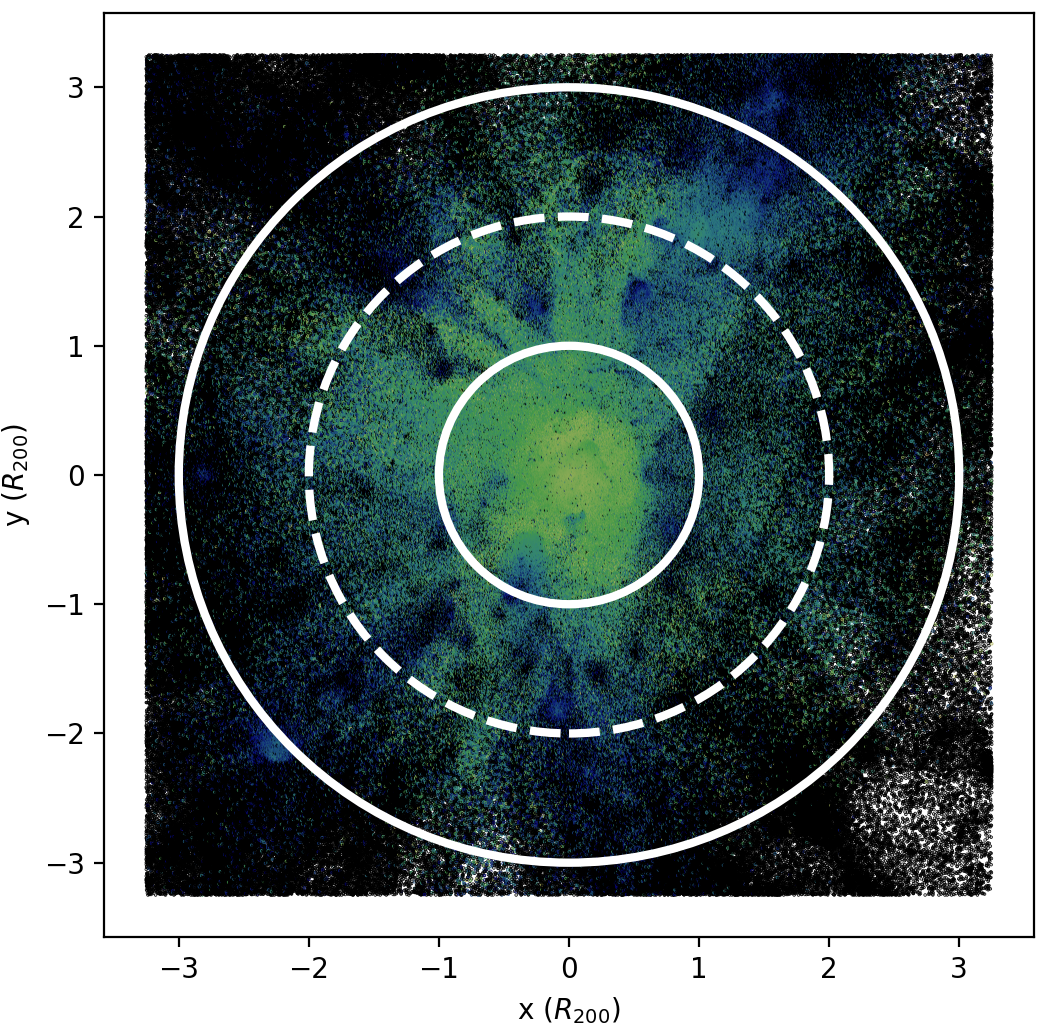}    
  }  
  \caption{A visual impression of Cluster 19 and its local environment. Here we
    show projected gas densities (upper panels) and temperatures (lower panels)
    in the fiducial non-radiative and full physics {\small G3-MUSIC}
    runs at $z$=0. Inner solid/dashed/outer solid circles indicate 1/2/3
    times $R_{200}$. In plotting maps of projected gas densities with
    {\small py-sphviewer} \citep{py-sphviewer}, smoothing lengths
    are estimated on a per particle basis before projection, allowing for
    fine detail in gas density to be discrened; temperature maps are
    constructed using the {\small matplotlib} hexbin function.
}
  \label{fig:visual_impression}
\end{figure*}

In Figure~\ref{fig:visual_impression}, we show projected gas density (upper
panels) and temperature maps (lower panel) for the {\small G3-MUSIC} run at
$z$=0. On the left, we show results for  the fiducial non-radiative
run, while on the right, we show one of its full physics counterparts.
Positions are plotted relative to $\vec{R}_{\rm cen}$ and normalised by $R_{200}$,
both evaluated for the particular run; we project from within a cube of
side $7\,R_{200}$, or equivalently out to a radius of $3.25\,R_{200}$. The
projected density maps were calcuated using the {\small py-sphviewer}
package \citep{py-sphviewer}, which computes smoothing lengths on a per
particle basis and so naturally adapts to regions of local density, allowing
fine structure to be discerned. For projected temperature maps, we use
{\small matplotlib}'s {\small hexbin} with a fine grid of $1536^2$.

Because the simulations were run with aligned parameters, we find excellent
agreement between the runs on intermediate-to-large scales; the structure of
the filamentary network, as traced by the gas, in which the cluster is
embedded is in very good agreement, and there is excellent 
correspondence between the positions of the more massive substructures
\citep[cf.][]{2016MNRAS.457.4063S}. The differences we see within the
core of the cluster have been studied previously
\citep[cf.][]{2016MNRAS.458.4052C}, and are driven by hydrodynamical shocks
and the physics of galaxy formation (i.e. cooling and feedback).

In contrast, we see clear differences between the runs on small-scales.
In the projected gas density maps, we see more high-density small-scale
structure in the full physics run than in the galaxy formation run
throughout the volume, as we would expect in the presence of radiative
cooling. However, in the projected temperature maps, we see more sharply
defined features in the non-radiative run - for example, the ridge
extending from upper-left to lower-right in the cluster core, as well as
several cool, dense knots with $R_{200}$ - than in the full physics run. We
shall return to these observations in the next section.

\section{Results}
\label{sec:results}
We now investigate the thermodynamical properties of the ICM in the outskirts
of the cluster at $z$=0. In particular, we focus on radial profiles of gas
density, temperature, and radial velocity in the regime
$0.3 \leq R/R_{\rm 200} \lesssim 3$, and investigate in detail the relationship
between structures evident in these profiles, which we interpret as arising
from shocks, and the orbital motions of substructures (\S~\ref{ssec:mergers})
and accretion from the cosmic web (\S~\ref{ssec:lss}). Because we have a
range of astrophysical codes and galaxy formation prescriptions, we can
estimate the degree of variation that arises. Our reference simulation is the
non-radiative {\small G3-MUSIC} run, and residuals (when shown) are with
respect to this run. All figures include data from both non-radiative
and full physics simulations, unless otherwise stated.

\subsection{Radial profiles of ICM gas properties}
\label{ssec:spherical_averages}

\begin{figure}
  \includegraphics[width=0.95\columnwidth]{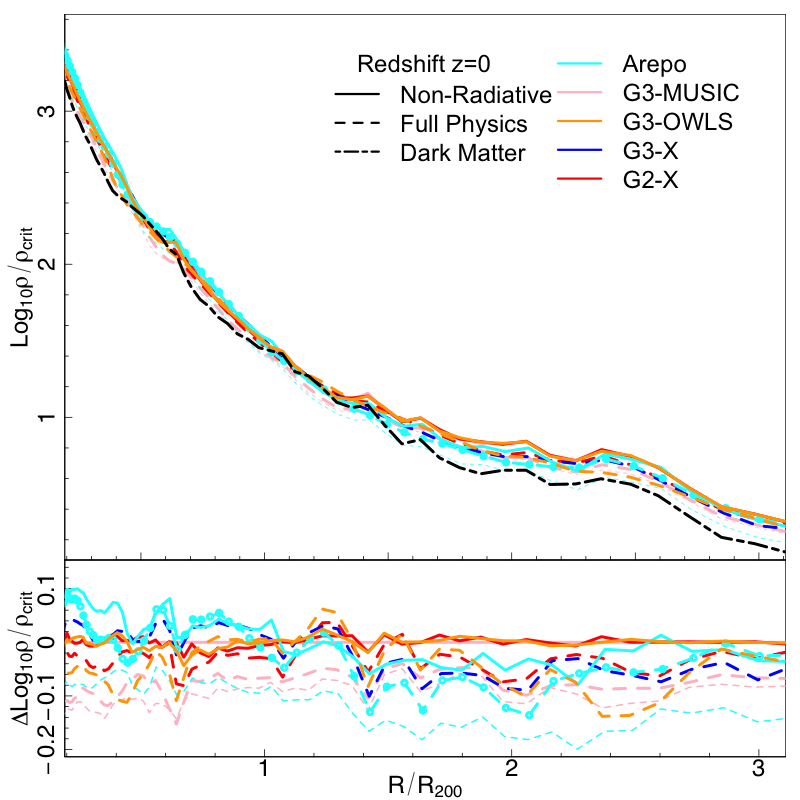}
  \caption{Spherically averaged gas density profiles at $z$=0. In the upper panel,
    we show the spherically averaged gas density profiles measured in the non-radiative
    runs (solid curves) and full physics runs (dashed curves). Here radii are normalised
    to the value of $R_{200}$ in the {\small G3-MUSIC} non-radiative run; densities are in
    units of $\rho_{\rm crit}$ at $z$=0, 2.7755$\times10^{11}\,h^2\,{\rm M}_{\odot}\,{\rm Mpc}^{-3}$.
    For reference, we show the heavy dotted-dashed curve is the dark matter density profile (as found in the dark
    matter only {\small G3-MUSIC} reference run) scaled by the cosmic baryon density,
    $\Omega_{\rm bar}$/$\Omega_{\rm m}$. In the lower panel, we show the residuals of each of
    these profiles with respect to our {\small G3-MUSIC} non-radiative reference run. Note that there are
    two sets of {\small Arepo} and {\small G3-MUSIC} full physics runs -
    the {\small Arepo-IL} run with AGN (heavy dashed curves with filled
    circles), the {\small Arepo-SH} run with stellar feedback but without AGN
    (light dashed curves), and the {\small G3-MUSIC-SH} and
    {\small G3-MUSIC-PS} runs (heavy and light dashed curves) with stellar
    feedback and without AGN.}
  \label{fig:density_profile_z0}
\end{figure}

\begin{figure}
  \includegraphics[width=0.95\columnwidth]{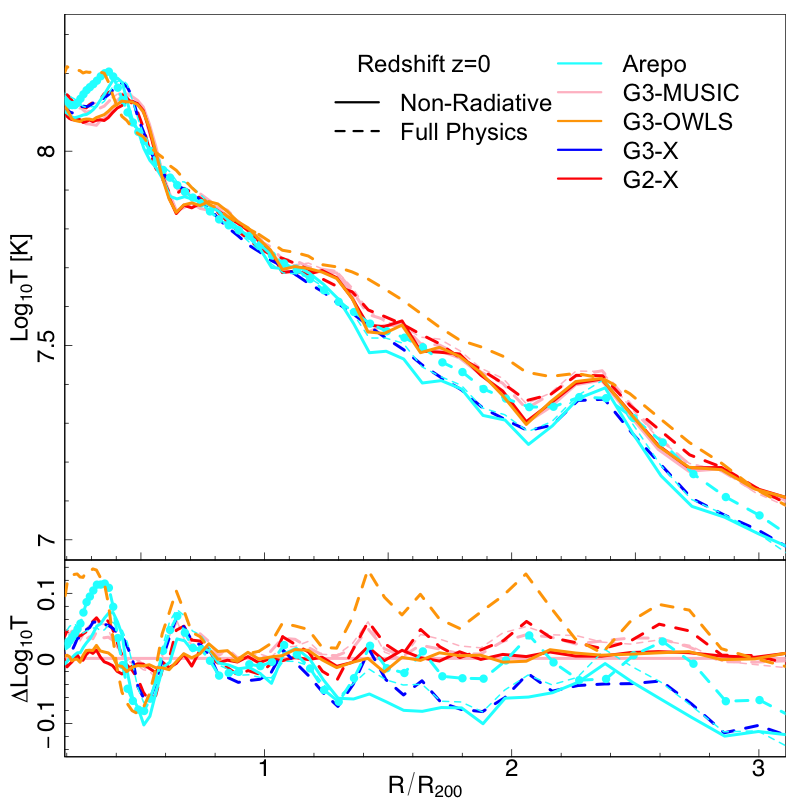}
  \caption{Spherically averaged gas temperature profiles at $z$=0. In the upper panel,
    we show the spherically averaged gas density profiles measured in the non-radiative
    runs (solid curves) and full physics runs (dashed curves). Radii are again normalised
    to $R_{200}$. In the lower panel, we show the residuals of each of these profiles with
    respect to the {\small G3-MUSIC} non-radiative run.
    We follow the same convention as in Figure~\ref{fig:density_profile_z0}
    to distinguish between the different full physics {\small Arepo} and
    {\small G3-MUSIC} runs.}
  \label{fig:temperature_profile_z0}
\end{figure}

We begin our analysis by considering spherically averaged gas density and
temperature profiles measured in the non-radiative (solid curves) and full
physics (long dashed curves) simulations in our sample. Here we estimate the
gas density at radius $R$ measured with respect to $\vec{R}_{\rm cen}$ as the
total gas mass (i.e. summation over gas particle masses) within a spherical
shell, divided by the volume of the shell; gas temperature is deduced from
the average, mass-weighted, specific internal energy of particles in the
shell, multiplied
by $3\mu\,m_p/2\,k_{\rm B}$, where $\mu$ is the mean molecular weight, $m_p$ is
the proton mass, $k_{\rm B}$ is the Boltzmann constant, and we have assumed
an adiabatic index of $\gamma=5/3$.

We show these spherically averaged profiles in
Figures~\ref{fig:density_profile_z0} (density) and
\ref{fig:temperature_profile_z0} (temperature). All radii are normalised
to $R_{200}$, the value in the fiducial {\small G3-MUSIC} non-radiative run;
densities are expressed in units of $\rho_{\rm crit}$, the critical density
at $z$=0; and temperatures are expressed in units of Kelvin. Lower panels
show residuals with respect to the fiducial. Note that there are
two curves for the {\small Arepo} and {\small G3-MUSIC} full physics
runs, corresponding to the two sets of galaxy formation models provided
- {\small Arepo-IL} (heavy dashed curve with filled circles)  which has
been run with AGN feedback, while the
{\small Arepo-SH} (light dashed curve), {\small G3-MUSIC-SH} (heavy dashed
curve), and
{\small G3-MUSIC-PS} (light dashed curve) runs
have been run without AGN feedback (see Appendix for more details).
For reference, in Figure~\ref{fig:density_profile_z0}, we overlay
the density profile of the dark matter only version of the cluster
(dotted-dashed curve), scaled by the universal baryon fraction,
$\Omega_{\rm bar}$/$\Omega_{\rm m}$. 

We find general agreement in the amplitudes and shapes of these spherically
averaged profiles measured in the different sets of runs, with differences
no greater than 0.1 dex ($\sim$25 per cent) over the radial range we consider,
as indicated by the residuals. The magnitude of the variations is substantially
smaller than seen for the central parts of the same cluster
\citep[see Figs.~8 and 9 in][]{2016MNRAS.tmp..582S}. There are small
enhancements in the spherically averaged gas density - for example, at
$\sim\!2\,R_{200}$ and $\sim\!0.7\,R_{200}$ - and they are broadly consistent
with the dark matter density profile scaled by the universal baryon fraction,
which provides a good
approximation to both the amplitude and shape of the gas density profiles,
which is consistent with the findings of earlier hydrodynamical simulations
\citep[e.g.,][]{2000ApJ...536..623L,2000MNRAS.317.1029P}
and the assumptions of analytical modeling
\citep[e.g.,][]{1998ApJ...497..555M,2001MNRAS.327.1353K}.

Enhancements in temperature are more pronounced. Within $R_{200}$, there is
notieceable bump at $\sim 0.5-0.6 R_{200}$, although its precise location and
shape varies between runs. We find that the non-radiative runs produce
temperature inhomogeneities that are more sharply pronounced than in their full
physics counterparts, which is consistent with our observations in
\S~\ref{ssec:visual_impression}. Although we expect that cooling in the full
physics runs should exacerbate existing inhomogeneities in temperature, the
net effect is to remove lower entropy, colder, denser, gas from the diffuse
phase, and weaken rather than enhance temperature inhomogeneities. We note
also that the variety of feedback processes that are active within a galaxy
cluster - for example, AGN and cosmic rays - will tend to smooth out
temperature inhomogeneities \citep[e.g.][]{2017MNRAS.467.3827P}.

\medskip

\begin{figure}
  \includegraphics[width=0.95\columnwidth]{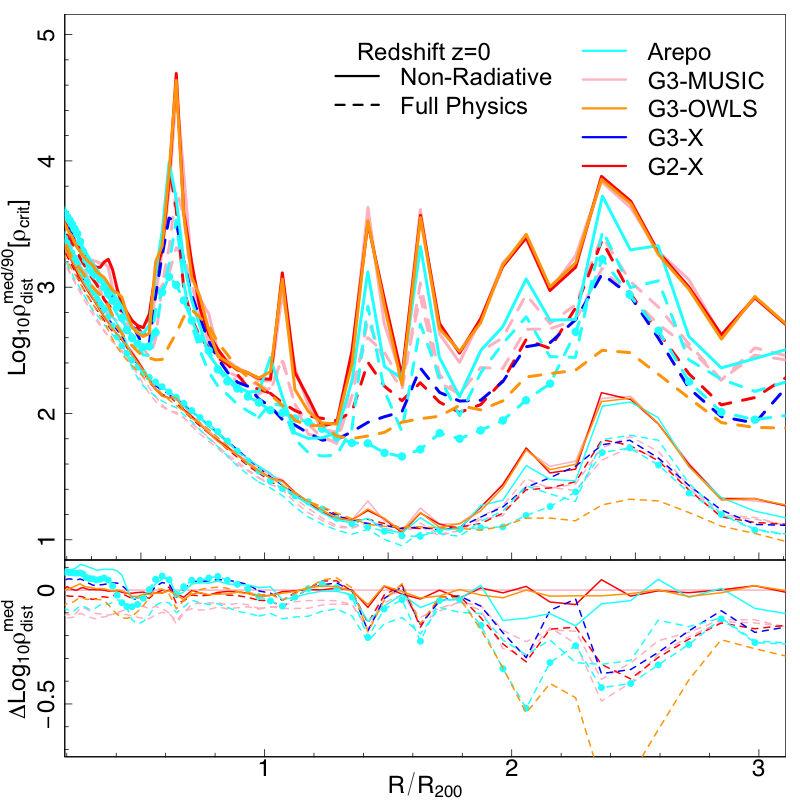}
  \caption{Radial profiles of median and $90^{\rm th}$ percentile gas density
    ($\rho_{\rm dist}^{\rm med}$, lower curves; and $\rho_{\rm dist}^{90}$, upper
    curves) at $z$=0, deduced from the distribution of gas densities within
    the spherical shells used in Figure~\ref{fig:density_profile_z0}. As
    before, we show results for non-radiative and full physics runs (solid
    and dashed curves respectively), and radii are normalised to $R_{200}$.
    In the lower panel, for clarity, we focus on the residuals of each of the
    median profiles with respect to the {\small G3-MUSIC} non-radiative run.
    Note that we
    distinguish between the two sets of full physics {\small Arepo} and
    {\small G3-MUSIC} runs as in Figure~\ref{fig:density_profile_z0}.}
  \label{fig:median_density_profile_z0}
\end{figure}

\begin{figure}
  \includegraphics[width=0.95\columnwidth]{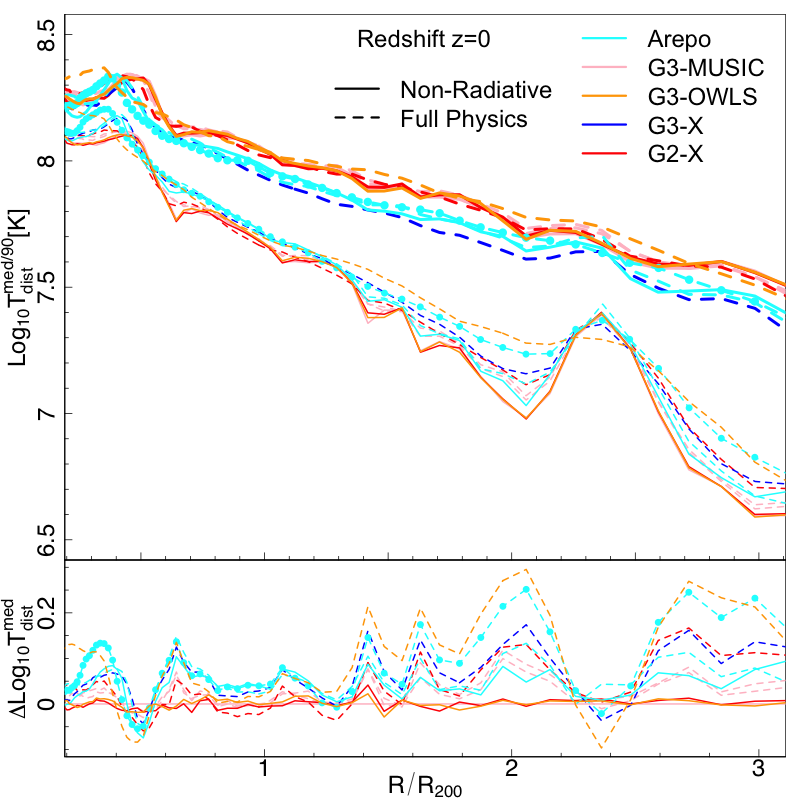}
  \caption{Radial profiles of median and $90^{\rm th}$ percentile gas temperature
    ($T_{\rm dist}^{\rm med}$, lower curves; and $T_{\rm dist}^{90}$, upper
    curves) at $z$=0, deduced from the distribution of internal energies within
    the spherical shells used in Figure~\ref{fig:density_profile_z0}.
    As in Figure~\ref{fig:median_density_profile_z0}, we consider only the
    residuals of the median profiles with respect to the {\small G3-MUSIC}
    non-radiative run, and we distinguish between the two sets of full physics
    {\small Arepo} and {\small G3-MUSIC} using the same convention.
  }
  \label{fig:median_temperature_profile_z0}
\end{figure}

An instructive alternative to spherically averaging profiles is to consider
measures of the distribution of values within each spherical shell, as explored
in \citet{2013MNRAS.428.3274Z} and subsequent papers. In
Figure~\ref{fig:median_density_profile_z0}, we show
how the median (lower curves) and $90^{\rm th}$ percentile values (upper curves)
of gas density, $\rho_{\rm dist}^{\rm med}$ and $\rho_{\rm dist}^{90}$, vary with
radius. Here the gas density is the intrinsic value
(for a particle or cell) tracked in the simulation. We find that the median
profile ($\rho_{\rm dist}^{\rm med}$) and the corresponding spherically averaged
profile in Figure~\ref{fig:density_profile_z0} are in very good agreement
within $R_{200}$, at the $5-10\%$ level.

The enhancement in $\rho_{\rm dist}^{\rm med}$ apparent at $R\simeq2\,R_{200}$
is consistent across codes and galaxy formation models, but its magnitude
differs between non-radiative and full physics runs by $\sim 30$ per cent. We
associate this with substructure infalling along the cosmic web, evident
in the upper right quadrants of the upper panels in
Figure~\ref{fig:visual_impression}. At $R \leq R_{200}$, pronounced
enhancements are apparent in $\rho_{\rm dist}^{90}$;
the locations of these peaks are broadly consistent between codes and
galaxy formation models, albeit with significant scatter in magnitude. We
find the stronger enhancements in the non-radiative runs (for the
reasons already discussed), but the larger scatter (by a factor of
$\sim 10$ in the enhancement at $\sim 0.6\,R_{200}$) in the full physics runs.
As noted by \citet{2013MNRAS.428.3274Z} and e.g. \citet{2016ApJ...833..227A},
these are naturally interpreted as signatures of substructure.

Figure~\ref{fig:median_temperature_profile_z0} shows that the density
enhancement in Figure~\ref{fig:median_density_profile_z0} at
$R\simeq2\,R_{200}$ has a corresponding temperature enhancement, which
is most easily discernible in the median profile ($T_{\rm dist}^{\rm med}$, lower
curves) and is most sharply defined in the non-radiative runs (again, for
the reasons already discussed). The enhancement
evident in the spherically averaged temperature profile at
$\sim 0.5-0.6 R_{200}$ is also apparent in $T_{\rm dist}^{\rm med}$
and $90^{th}$
percentile temperatures profile ($T_{\rm dist}^{90}$, upper curves), while smaller
features are evident at the locations of the density enhancements in
Figure~\ref{fig:median_density_profile_z0}. The sharpness of these
features varies systematically between non-radiative runs, which are
in very good agreement with each other, and the full physics runs, which tend
to produce less sharply defined features than in the non-radiative runs.

We have used snapshots finely spaced over the redshift range $z$=0.5 to $z$=0
to verify that the temperature enhancements within $R_{200}$ evident in these
radial profiles occur frequently during the cluster's assembly. This strengthens
our assetion that they are transient features whose position moves
to smaller radii over a dynamical time, consistent with the motions of
substructures whose passage shocks the ICM.
In contrast, the feature at $\sim(2-3)\,R_{200}$ identified at
$z$=0 varies in amplitude but has remained at approximately $\sim(2-3)\,R_{200}$
since $z$=0.5, which we argue is associated with gas inflow and substructure
motion in the cosmic web \citep[see, e.g.,][]{2010ApJ...721.1105B}.

We note that \citet{2016ApJ...833..227A} considered the mean mass-weighted
temperature of the ICM at larger radius (defined relative to $R_{500}$, the radius enclosing a mean density of 500 times critical density)
of an ensemble average of clusters and
found that the presence of
substructure tended to reduce it compared to the temperature when substructure
was excluded, as a result of the cooler gas associated with substructures
bringing down the average temperature. This is consistent with our results
- we consider the temperature distribution of all gas within the shell, and so
the $90^{\rm th}$ percentiles will include the contributions of shocked gas,
some of which can be associated with the $90^{\rm th}$ percentiles in density.
We have verified that excluding overdense gas leads to an increase in the
mass weighted average temperature within shells, albeit our result is noisier
than that of \citet{2016ApJ...833..227A}.

\begin{figure}
  \includegraphics[width=0.95\columnwidth]{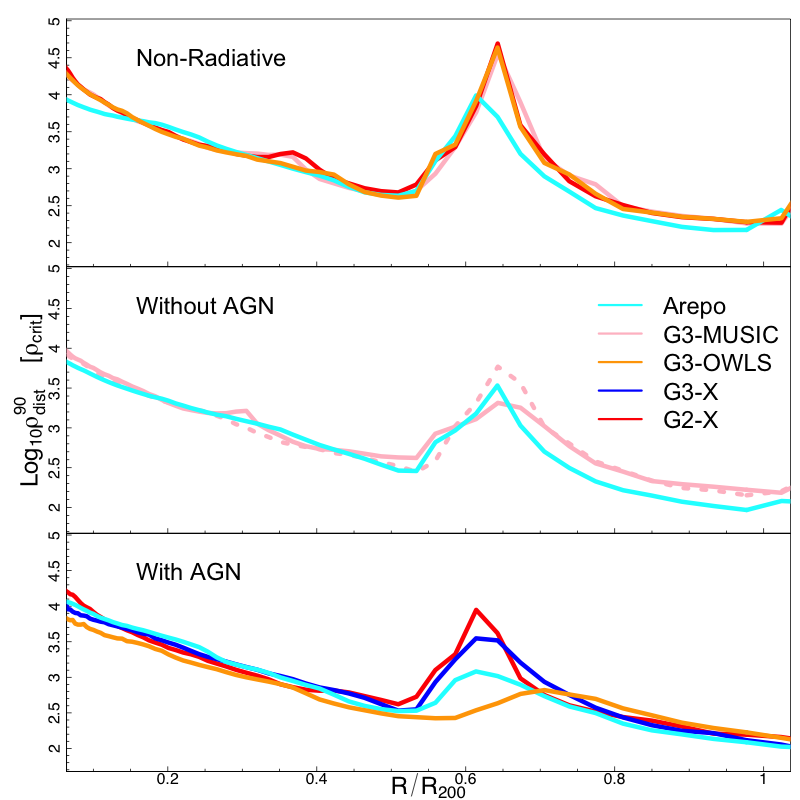}
  \includegraphics[width=0.95\columnwidth]{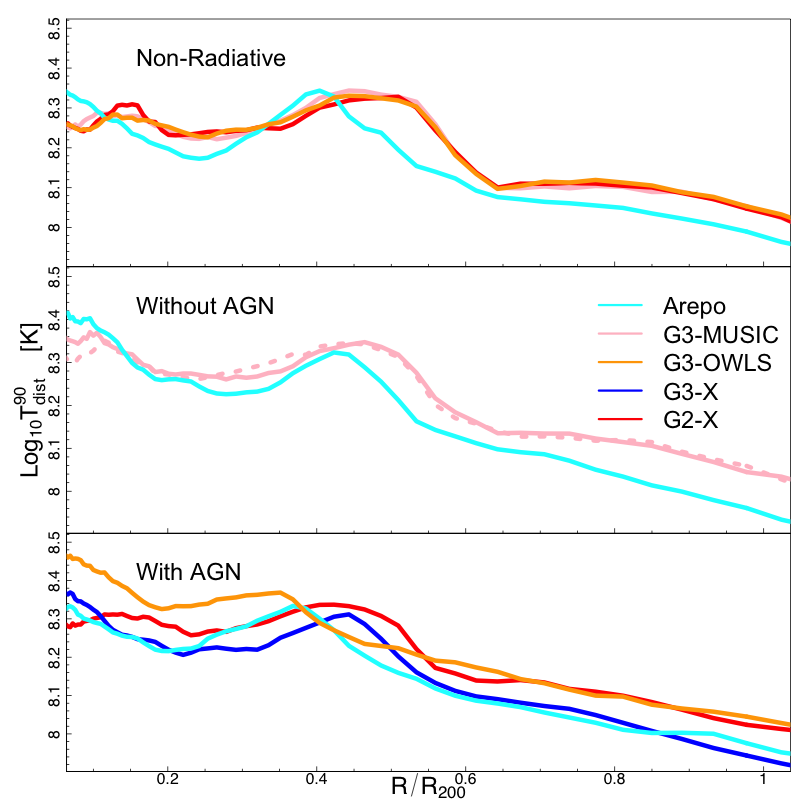}
  \caption{Radial profiles of the $90^{\rm th}$ percentile gas density (top
    panels) and temperature (bottom panels), computed as in
    Figures~\ref{fig:median_density_profile_z0} and
    ~\ref{fig:median_temperature_profile_z0}. We distinguish between
    non-radiative runs (upper panel) and full physics runs ``Without AGN''
    and ``With AGN'' (middle and lower panels, respectively). As before,
    there are two full physics {\small G3-MUSIC} runs without AGN -
    {\small G3-MUSIC-SH} and {\small G3-MUSIC-PS} - which we distinguish by
    heavy and light dashed curves, while there while there is one
    {\small Arepo} full physics run with AGN and one
    without.}
  \label{fig:profiles_by_physics}
\end{figure}

\begin{figure}
  \includegraphics[width=0.95\columnwidth]{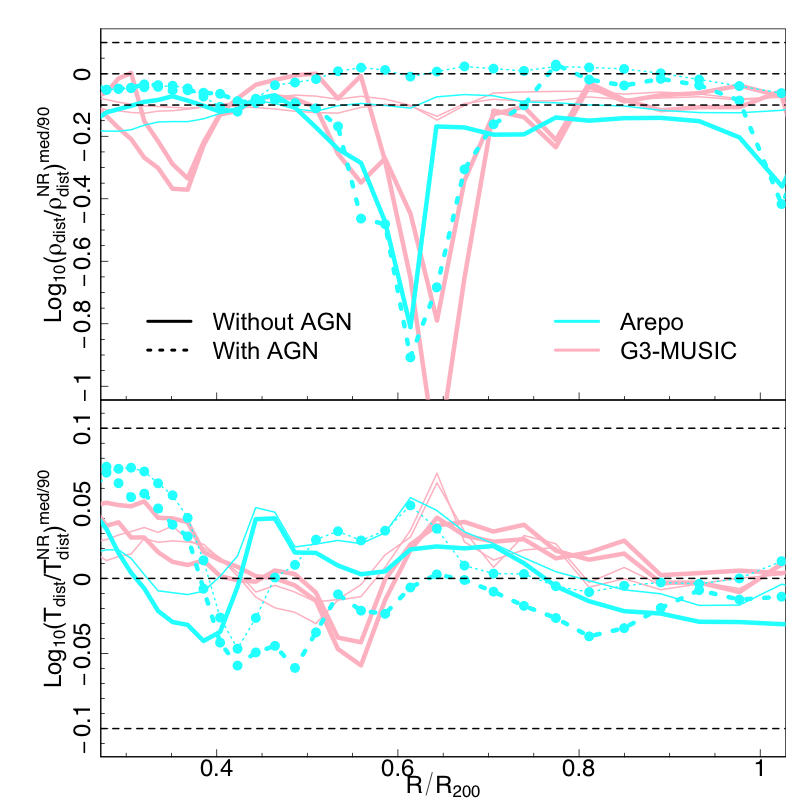}
  \caption{Comparison of density and temperature profiles in the
    {\small Arepo} and {\small G3-MUSIC} simulations, for which we
    have multiple full physics models, as in
    Figures~\ref{fig:median_density_profile_z0} and
    ~\ref{fig:median_temperature_profile_z0}. These profiles are
    scaled by the density and temperature profiles in the corresponding
    non-radiative runs. Light (heavy) curves are based on the medians
    ($90^{\rm th}$ percentiles), while we distinguish between runs
    ``Without AGN'' (the two {\small G3-MUSIC} runs and one of the
    {\small Arepo} runs) and ``With AGN'' (the other {\small Arepo} run)
    by using solid and dashed curves; for emphasis, we also overlay filled
    circles on the ``With AGN'' curves. Horizontal dashed lines delineate
    $\pm 25\%$.}
  \label{fig:model_comparison}
\end{figure}

\medskip

How might different physical processes modelled in the simulations impact
radial density and temperature profiles? In
Figure~\ref{fig:profiles_by_physics}, we
investigate how these profiles are influenced by the physical processes
modelled in the simulation, focusing on the $90^{\rm th}$ percentile values
of the density and temperature distributions within spherical shells where
we expect differences between the simulations to be most marked. Here we
group the simulations into (1) non-radiative runs, and full physics runs
(2) with, and (3) without AGN. Note that we zoom in on the radial range
$0.1 \leq R/R_{200} \leq 1$ to assess the extent to which the physics of
galaxy formation - particularly AGN feedback - might affect the ICM outside
of where the central cluster galaxy resides.

\medskip

\noindent{\bf{(1) Non-radiative Case:}} We see a sharply peaked density
enhancement at $\sim 0.6 R_{200}$, evident in all but one of the runs, which
is most pronounced in the non-radiative runs. There is excellent agreement
between the classic SPH codes ({\small G3-MUSIC}, {\small G2-X}, and
{\small G3-OWLS}) - unsurprising, given that they are variants of GADGET2/3
\citep{2005MNRAS.364.1105S} with the same gravity solver and the treatment
of SPH - while {\small Arepo} underpredicts the magnitude of the density
enhancement in comparison, but otherwise tracks the predictions of the classic
SPH codes.
Similarly, there is excellent consistency between the temperature profiles
predicted by the classic SPH codes, which show an enhancement at
$\sim 0.5R_{200}$, whereas {\small Arepo} underpredicts the temperature
across most of the radial
range, while the enhancement is offset to $\sim 0.4R_{200}$. Note that the
{\small Arepo} density and temperature profiles begin to peel away from their
classic SPH counterparts at $\sim 0.1R_{200}$, as we expect because of the
difference in hydrodynamics solvers \citep[e.g.][]{2016MNRAS.457.4063S}.

\medskip

\noindent{\bf{(2) Full Physics without AGN Case:}} The trends evident in the
non-radiative runs are also apparent in the full physics runs without AGN
- {\small Arepo-SH}, {\small G3-MUSIC-SH}, and {\small G3-MUSIC-PS}.
The amplitude of the density enhancement at $\sim 0.6 R_{200}$
is reduced relative to in the non-radiative runs, and we see from the
{\small G3-MUSIC} that the assumed stellar feedback model has an impact.

\medskip

\noindent{\bf{(3) Full Physics with AGN Case:}} The amplitude of the density
enhancement at $\sim 0.6 R_{200}$ shows considerable variation between the
different full physics runs that implement AGN feedback, with the two classic
SPH codes {\small G2-X} and {\small G3-OWLS} bracketing the range of variation
- the enhancement is peaked in {\small G2-X}, whereas it is effectively
smoothed away in {\small G3-OWLS}. A similar range of variation is evident
in the temperature profiles, with the amplitude, sharpness, and, in particular,
the location of the peak of the temperature enhancement at $\sim 0.5R_{200}$;
for example, the enhancement in the {\small G3-OWLS} run peaks at
$\sim 0.35 R_{200}$, compared to $\sim 0.45 R_{200}$ in the {\small G2-X} and
{\small G3-X} runs. Interestingly, the classic SPH runs predict systematically
higher temperatures across most of the radial range when compared to the
modern SPH run with {\small G3-X} and {\small Arepo}.

\medskip

These observations show that the physical processes modelled in a
simulation can influence the ICM density and temperature radial profiles
at larger radii, but the most significant differences are associated with
the density and temperature enhancement associated with substructure. Where
we see systematic differences - for example, in the temperature profiles in the
``With AGN'' runs - these can be attributed to expected differences between
classic SPH on the one hand, and modern SPH and {\small Arepo} on the other.

\medskip

To isolate the particular effect of feedback on the ICM density and
temperature radial profiles, in Figure~\ref{fig:model_comparison} we look
at the profiles in the {\small Arepo-SH}
and {\small Arepo-IL} runs, which are run without and with AGN feedback, and in
the {\small G3-MUSIC-SH} and {\small G3-MUSIC-PS}, which are run without
AGN feedback but utilise two different stellar feedback models
\citep[cf.][]{2003MNRAS.339..289S,2011MNRAS.410.2625P}. We show logarithmic
differences with respect to the non-radiative counterparts, and plot
medians (light curves) and $90^{\rm th}$ percentiles (heavy curves)
in the radial range $0.3 \leq R/R_{200} \leq 1$; models with AGN are indicated
by dotted curves overlaid with filled circles. The light dashed horizontal
lines indicate $\pm 25\%$ with respect to the non-radiative run.
As we would expect, the logarithmic difference is dominated by the density
enhancement at $\sim 0.6R_{200}$; logarithmic differences in the temperature
profiles are small ($\lesssim 0.05$ dex), while the corresponding differences
in density are predominantly $\lesssim 0.1$ dex, although these increase
where we see enhancements that we associated with substructure.

\bigskip

\begin{figure}
  \includegraphics[width=0.95\columnwidth]{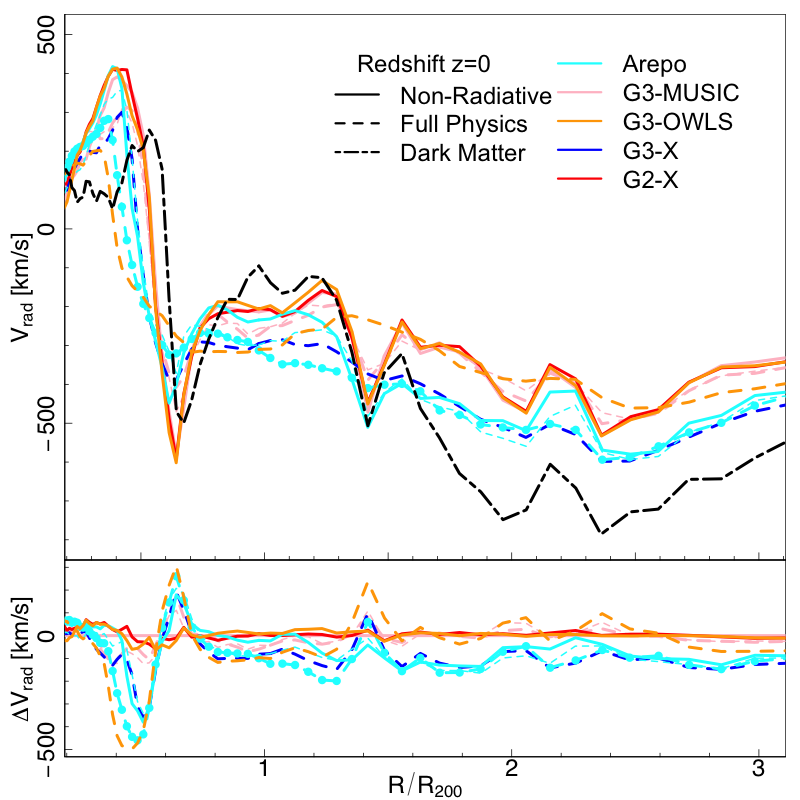}
  \caption{Spherically averaged radial velocity profiles at $z$=0. In the upper panel,
    we show the spherically averaged radial velocity profiles measured in the non-radiative
    runs (solid curves) and full physics runs (dashed curves). Velocities are expressed
    in units of km/s, while we normalise radii by $R_{200}$ as before.
    For reference, we show by the heavy dotted-dashed curve the dark matter radial velocity profile
    as found in the {\small G3-MUSIC} non-radiative run. In the lower panel,
    we show the residuals of each of these profiles with respect to the {\small G3-MUSIC}
    non-radiative run again.}
  \label{fig:radial_velocity_profile_z0}
\end{figure}

In Figure~\ref{fig:radial_velocity_profile_z0}, we shift focus to the radial
velocity
profile of gas within the cluster, reasoning that features in the temperature
profile are a response to gravitational and hydrodynamical influences; namely,
shocks. Here we show the variation of spherically averaged radial velocity of
gas with radius,
with radii normalised to $R_{200}$ while radial velocities are in units of km/s;
$v_{\rm rad}\!<\!0\,(>\!0)$ implies inflow (outflow). For a system in
hydrodynamic equilibrium, we would expect $\langle v_{\rm rad}\rangle\simeq 0$,
but instead we see a pronounced dip to $\sim -500\,{\rm km}{\rm s}^{-1}$ at
$\sim 0.7\,R_{200}$, before a sharp rise to $\sim 400\,{\rm km}{\rm s}^{-1}$
peaking at $\sim 0.5\,R_{200}$. There is appreciable variation between the
runs, but the trend is systematic.

Interestingly, we see a feature in the dark matter radial velocity (heavy
dotted-dashed curve), similar to the feature evident in the gas radial
velocity, albeit smaller in amplitude and displaced outwards in
radius. At radii $R \gtrsim r_{200}$, we see substantial differences between
runs of $\sim\!200$ km/s. It is noteworthy that the unstructured mesh code,
{\small Arepo}, and the modern SPH code, {\small G3-X}, predict
systematically more negative radial velocities than the other codes, which
are based on classic SPH, which are in broad agreement with one another.
Strikingly, the dark matter radial velocity profile more negative than the
gas radial velocity profiles by $\sim 200-400$ km/s, but there are common
features across the various profiles (for example, the bump at
$\sim\!2\!R_{200}$).

\medskip

\begin{figure}
  \includegraphics[width=0.95\columnwidth]{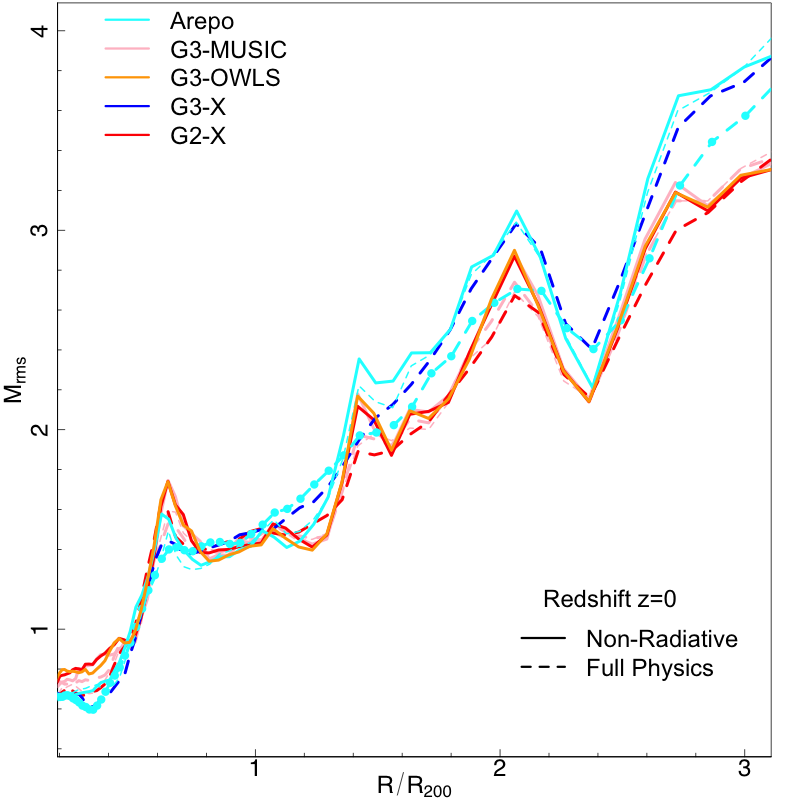}
  \caption{Spherically averaged radial profiles of the root-mean-square Mach number,
    $M_{\rm rms}$, at $z$=0. As before, we show results from the non-radiative
    runs (solid curves) and full physics runs (dashed curves), and radii are
    normalised by $R_{200}$ as before.}
  \label{fig:radial_mach_number_profile_z0}
\end{figure}

In Figure~\ref{fig:radial_mach_number_profile_z0}, we provide an estimate for the
strength of the shocks in the cluster gas by calculating the spherically averaged
Mach number of gas as a function of cluster-centric distance.
There are several approaches that we could take to estimating where and when shocks
occur, which include tracking spatially localised jumps in temperature and/or 
velocity \citep[e.g.][]{2009MNRAS.395.1333V}, which are relatively straightforward
in mesh-based codes where there are well-defined boundaries between cells; or 
tracking jumps in entropy or internal energy over time \citep{2003ApJ...585..128K},
which is particularly straightforward in particle-based codes. However, we follow
\citet{2019MNRAS.484.4881M} and use the root-mean-square (rms) Mach number,
\begin{equation}
  M_{\rm rms}=\frac{\sqrt{\langle{v^2}\rangle}}{c_s},
\end{equation}
where $\langle{v^2}\rangle$ is the rms velocity computed for all gas
elements (particles or cells) within a spherical shell, and $c_s$ is the sound speed,
which we compute directly from the internal energies.

As we would expect, the locations of features in the spherically averaged temperature and
radial profiles correlate with features in the rms Mach number radial profile. The
magnitude of $M_{\rm rms}$ is larger in the moving-mesh {\small Arepo} runs - both the
non-radiative and full physics variants - and in the modern SPH {\small G3-X} run, compared
to the classic SPH runs, in the regime of stronger
shocks, where $M_{\rm rms}\gtrsim 2$ and $R\gtrsim{1.5}R_{200}$, by between 10\% and 25\%.
The trend is for the average rms Mach number to increase with cluster-centric distance,
with values of $M_{\rm rms}\lesssim{1}$ within $0.5\,R_{200}$ and $M_{\rm rms}\lesssim{1.8}$
within $R_{200}$, rising to $M_{\rm rms}\lesssim{3-4}$ at $3\,R_{200}$. Note, however, that
these numbers are indicative - if, instead, we measure median ($90^{\rm th}$ percentiles)
values of $\langle{v^2}\rangle$ within each shell, we find $M_{\rm rms}\lesssim{2} (3)$
within $R_{200}$ and $M_{\rm rms}\lesssim{5} (50)$ at $3\,R_{200}$. Nevertheless, the values
are broadly consistent with reported values in the literature. For example,
\citet{2009MNRAS.395.1333V} note that ``internal shocks'' within the virialised region of a
cluster tend to have Mach numbers of $M\lesssim 3$, with occasional spikes as a result of
mergers, while ``external shocks'' tend to be much stronger and larger ($M\gg{10}$), arising
from accretion onto filaments and sheets \citep[see also][]{2001ApJ...562..233M}.

\medskip

We have demonstrated that enhancements in gas density and temperature profiles
within $R_{200}$, as well as features in gas and dark matter radial velocity
profiles, plausibly correspond to substructures evident in projected gas
density and temperature maps (Figure~\ref{fig:visual_impression}).
We now investigate how substructure gives rise to the features in detail.

\medskip

\subsection{The impact of substructure dynamics on the ICM}
\label{ssec:mergers}

So far we have identified the imprint of substructure on the radial profiles
of the ICM gas density, temperature, and radial velocity; we now shift our
focus to the substructure population and investigate how its dynamical effects
impact the ICM.

\medskip
\begin{figure}
  \includegraphics[width=0.49\textwidth, trim=5.0cm 7.0cm 4.0cm 9.0cm, clip=true]{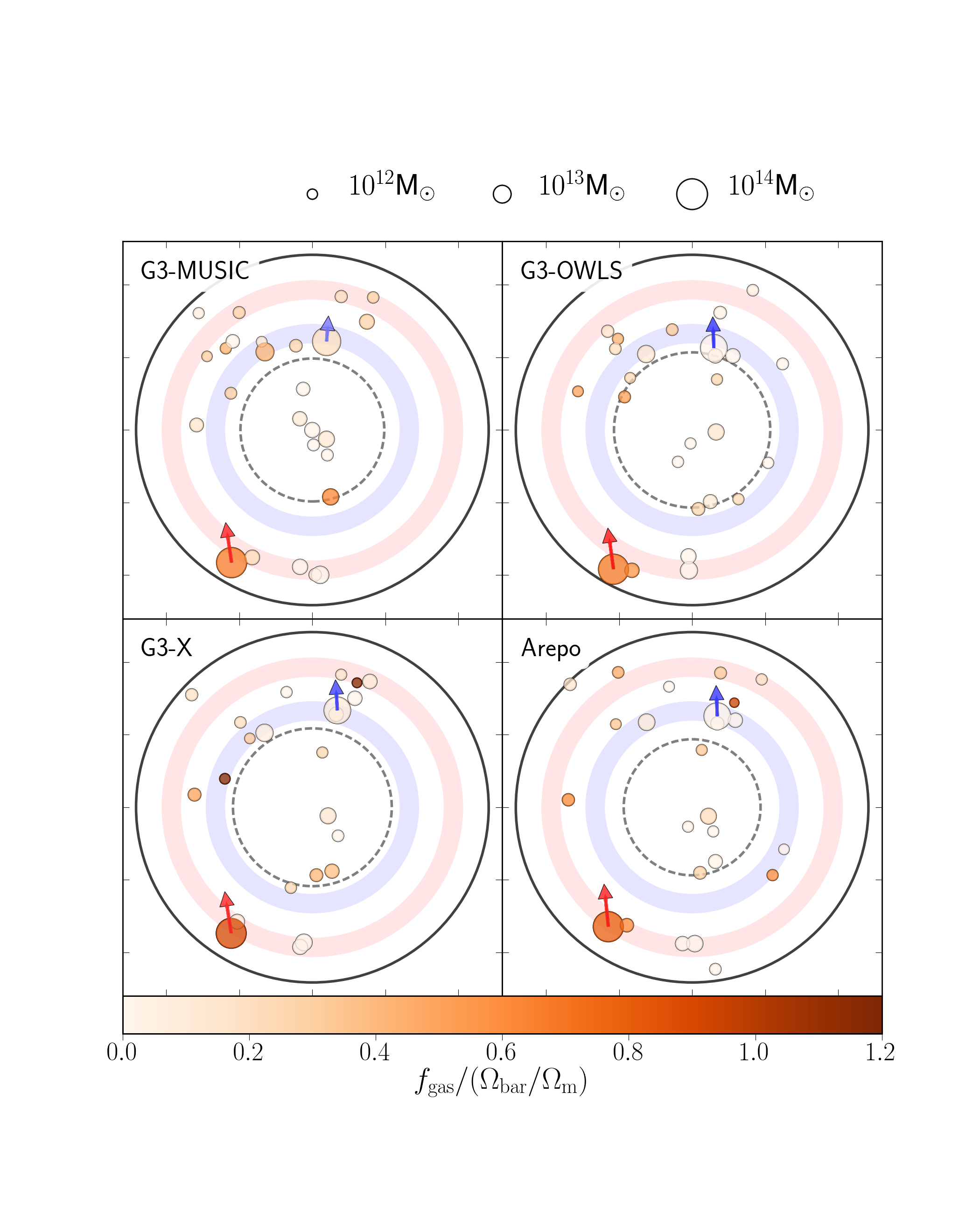}
  \caption{Projected spatial distribution of haloes within $1.25R_{200}$ of
    the cluster. We include only
    those haloes with masses $\geq 10^{12}h^{-1} {\rm M}_{\odot}$. Symbol size
    and colour scale with mass and gas fraction, with lighter hues indicating gas
    paucity and dark hues indicating gas richness. The virial radius $R_{200}$ is
    represented by the solid circle, the radius at which the underlying cluster
    halo circular velocity profile reaches its maximum by the dashed circle, while
    the shaded regions indicate the inner and outer boundaries of the temperature
    and radial velocity profile features.}
  \label{fig:clus19haloraddistrib}
\end{figure}

\begin{figure*}
  \centering{
  \includegraphics[height=6cm, trim=6.0cm 2.0cm 17.0cm 2.0cm, clip=true]{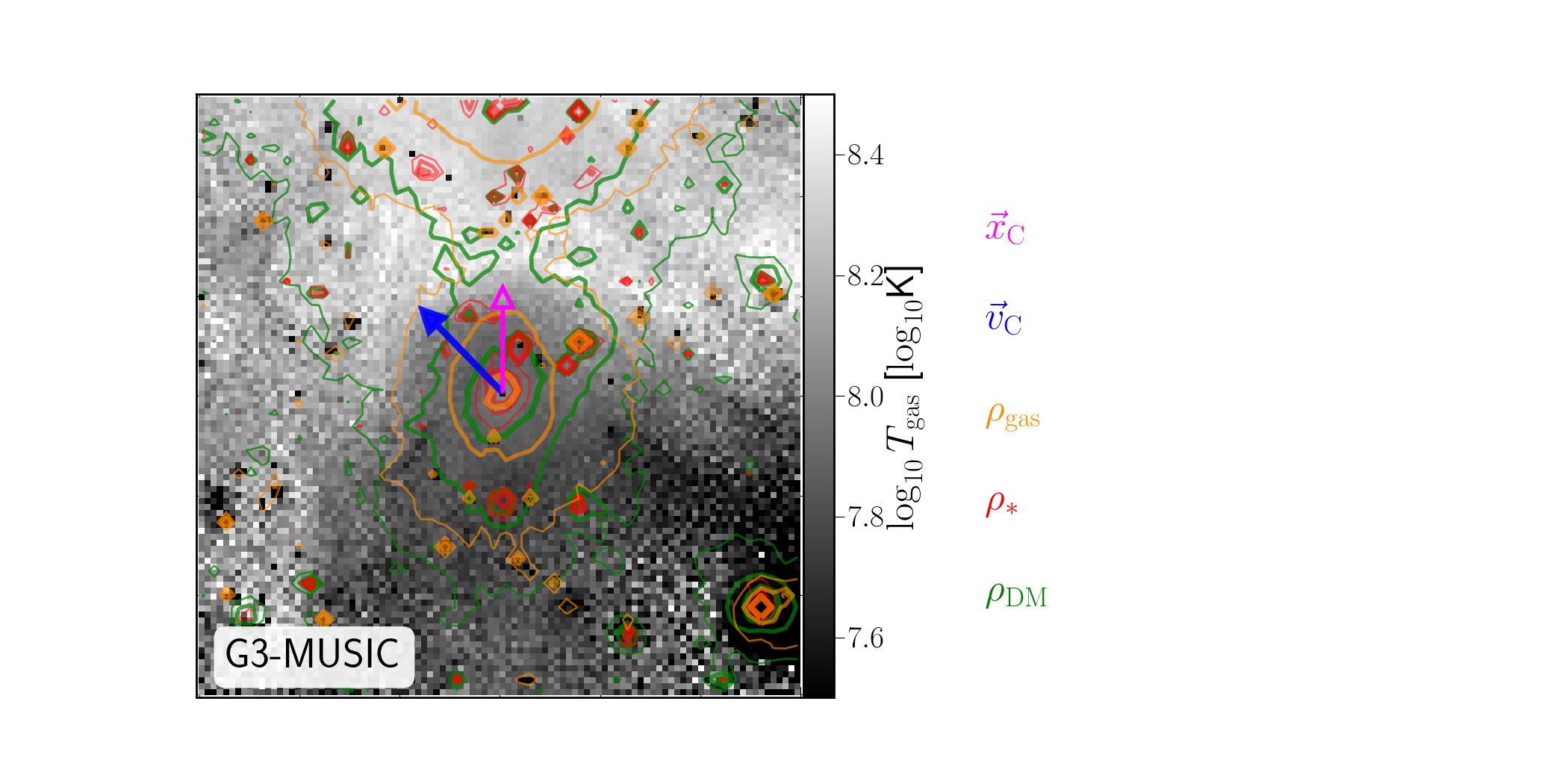}}
  \includegraphics[height=6cm, trim=6.0cm 2.0cm 21.0cm 2.0cm, clip=true]{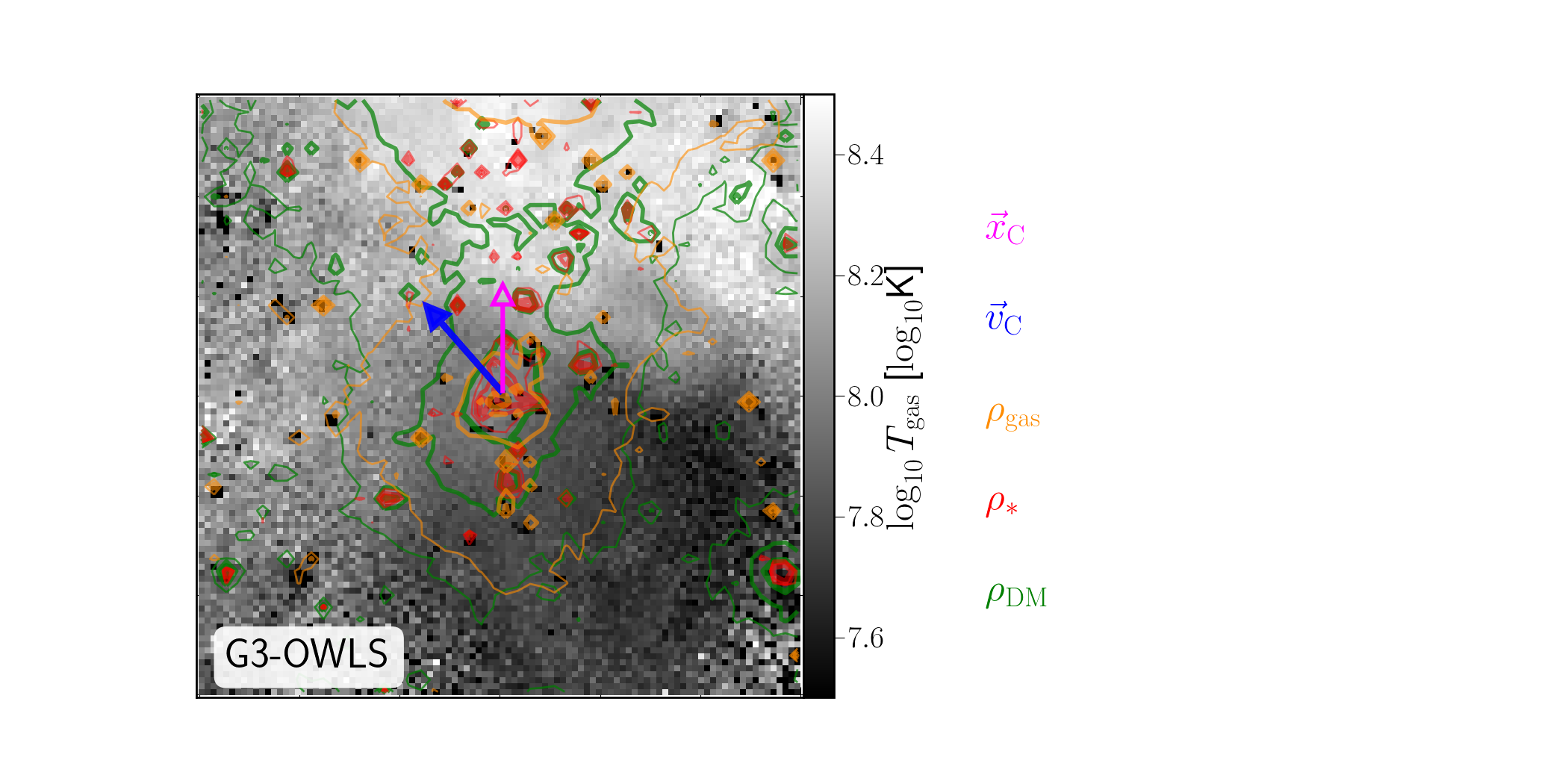}
  \centering{
  \includegraphics[height=6cm, trim=6.0cm 2.0cm 17.0cm 2.0cm, clip=true]{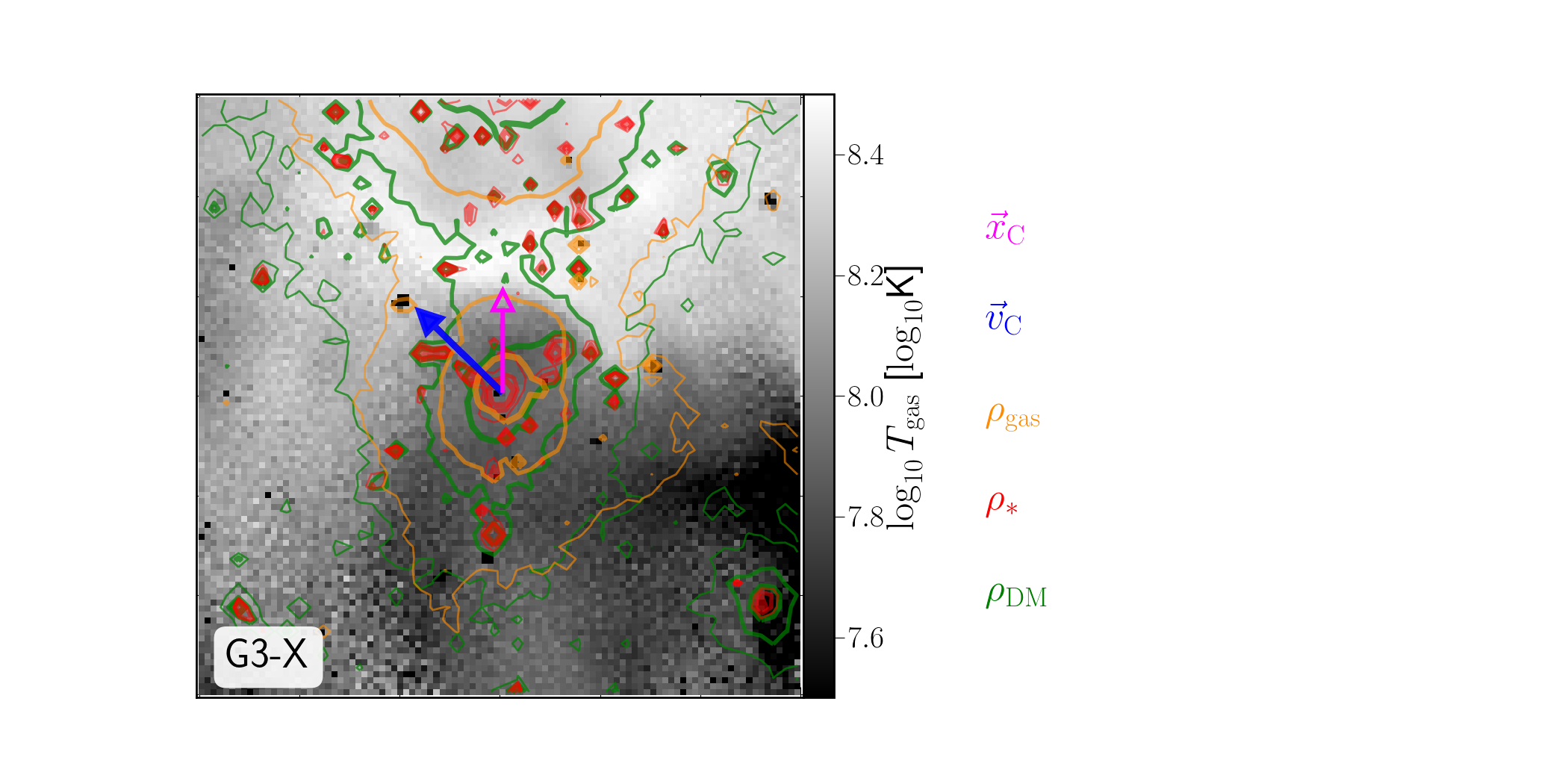}}
  \includegraphics[height=6cm, trim=6.0cm 2.0cm 21.0cm 2.0cm, clip=true]{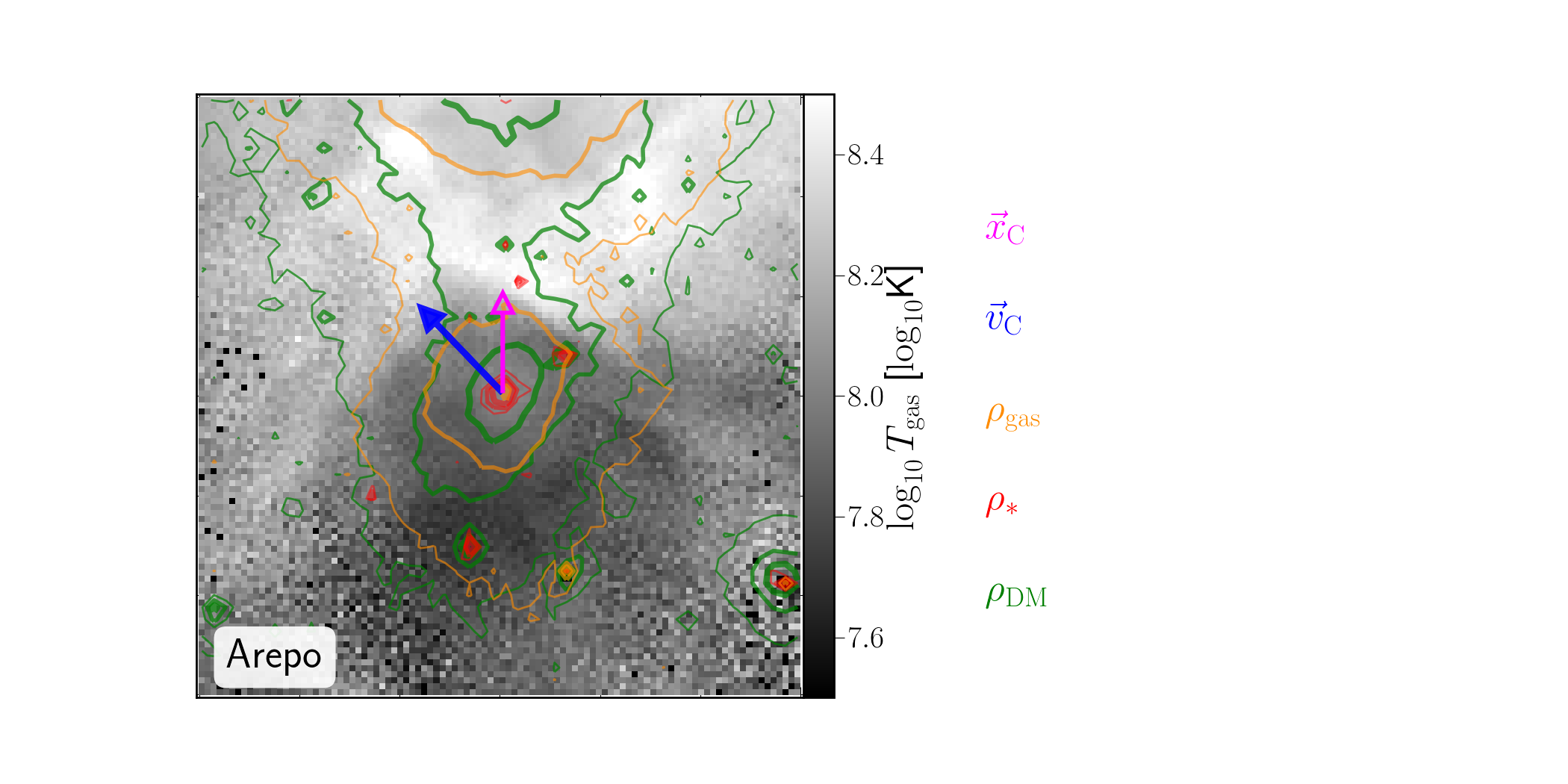}
  \caption{Infalling group shock in a subset of the full physics runs; from top right to bottom left,
    {\small G3-MUSIC}, {\small G3-OWLS}, {\small G3-X}, and {\small Arepo}. Here we
    zoom in on a region $2.1h^{-1} {\rm Mpc}$ wide and $1.1h^{-1} {\rm Mpc}$ thick, centred on and in the orbital
    plane of the galaxy group that generates the shock that gives rise to the feature in
    the temperature profile at $0.9R_{200}$.
    We show the mean logarithmic temperature distribution along the line of sight in this
    slab in grayscale, where each pixel is $20$ kpc in size. We show also contour maps of
    the average dark matter density in green; the average gas density in orange; and the average
    stellar density in red. The galaxy group's direction of motion relative to the cluster's
    centre-of-mass is shown by a blue arrow, while the direction towards the cluster centre is shown by
    by a magenta arrow.}
  \label{fig:clus19shockregions-fp}
\end{figure*}

In Figure~\ref{fig:clus19haloraddistrib} we show the projected spatial distribution of haloes of mass
$M_{\rm 200}\geq10^{12}h^{-1} {\rm M}_{\odot}$ and within $1.25 R_{\rm 200}$ of the cluster's centre in the
full-physics simulations. Symbol size is proportional to halo mass, while symbol colour indicates gas
fraction. The solid outermost circle represents the cluster's radius $R_{\rm 200}$, while the dashed
circle indicates the position of the peak of the circular velocity profile. The shaded regions mark the
outer and inner bounds of the features identified in the spherically averaged
temperature and radial velocity profiles (cf. Figures~\ref{fig:temperature_profile_z0} and~\ref{fig:radial_velocity_profile_z0}).

The most striking feature of Figure~\ref{fig:clus19haloraddistrib} is the
presence of two distinct massive substructures whose radial distances coincide
with features in the gas radial velocity profile. As a case study, we
investigate these substructures in more detail. Their trajectories are
consistent with the trends we see in the radial velocity profiles - the outer,
gas-rich substructure is infalling with a velocity $\sim -1700$~km/s, whereas the
inner, gas-poor substructure is moving outwards with a
velocity $\sim 1500$~km/s - although their velocities are much higher than the spherically averaged velocities at
these radii, and supersonic. Their velocity vectors are relatively well aligned, with an angle between them of
$\sim$$20^\circ$, and their direction of motion aligns with the orientation of the large-scale filaments within
which the cluster is embedded (see further discussion in \S~\ref{ssec:lss}).

These results are consistent with non-radiative and dark matter only versions
of these simulations\footnote{See also Fig.1 (DM) and Fig.5 (gas) in \citet{2016MNRAS.457.4063S} for a visualisation of the
  non-radiative runs as well as Figs.4 \& 6 in \citet{2016MNRAS.tmp..582S} for the counterparts
  in the full physics runs.}, which we check by cross-matching halos identified in the full physics
run with those in the non-radiative and dark matter only runs, and confirm excellent agreement
between the spatial and kinematic distributions. Differences that we note in the dark matter only run --
that the counterpart to the outgoing halo is at a slightly larger cluster-centric distance than in
the full physics and non-radiative runs -- can be readily understood as a consequence of the change
in bound mass arising from the stripping of its gas content.

\medskip

In Figure~\ref{fig:clus19shockregions-fp}, we make explicit how the infalling substructure, highlighted
in Figure~\ref{fig:clus19haloraddistrib}, impacts the ICM. Recall that
this infalling substructure is gas-rich ($f_g\sim(0.7-0.9)\Omega_b/\Omega_m$), and it has a large radial
velocity ($\sim$-1700km/s) relative to the cluster; this velocity is significantly larger than
the motions of cluster particles in the same region -- for dark matter,
$\left\langle v_r \right\rangle\sim -150$~km/s with dispersions of $\sim1000$~km/s; for gas
$\left\langle v_r \right\rangle\sim -250$~km/s, with dispersions of $\sim600$~km/s. We plot
the projected gas distribution, weighted by temperature, in a region $2.1h^{-1} {\rm Mpc}$ wide
and $1.1h^{-1} {\rm Mpc}$ thick, centred on and in the orbital plane of the substructure. Contours indicate
dark matter, gas, and stellar density associated with the substructure, while arrows show the direction
to the cluster centre ($\vec{x}_{\rm C}$) and the direction of motion of the substructure relative to the
cluster centre ($\vec{v}_{\rm C}$).

Figure~\ref{fig:clus19shockregions-fp} shows that the infalling substructure driving a shock into the ICM - there is evidence for shocked gas in advance of the substructure and trailing gas
in its wake in all of the runs. The temperature enhancement associated with the
shock is more pronounced in the direction of the cluster centre, towards which the density gradient
is higher, than in the direction of motion\footnote{Code-to-code variations are evident in the structure of the ICM - the classic
  SPH codes ({\small G3-MUSIC},{\small G3-OWLS}) show more small-scale variations
  than either the modern SPH codes ({\small G3-X}) or the moving-mesh code
  ({\small Arepo}), and the differences are particularly striking when one
  compares {\small G3-MUSIC}, which contains several cool, dense knots within
  this region (lower left), with {\small G3-X}, where such knots are absent.
  These small-scale variations are especially pronounced in the non-radiative
  runs, consistent with previous studies
  \citep[see, e.g., Figure 12 in][]{2014MNRAS.440.3243P},
  but the impact of these variations on our conclusions are quantitative, not
  qualitative.}.

\medskip
\begin{figure}
  \includegraphics[width=0.49\textwidth, trim=0.0cm 0.0cm 0.0cm 0.0cm, clip=true]{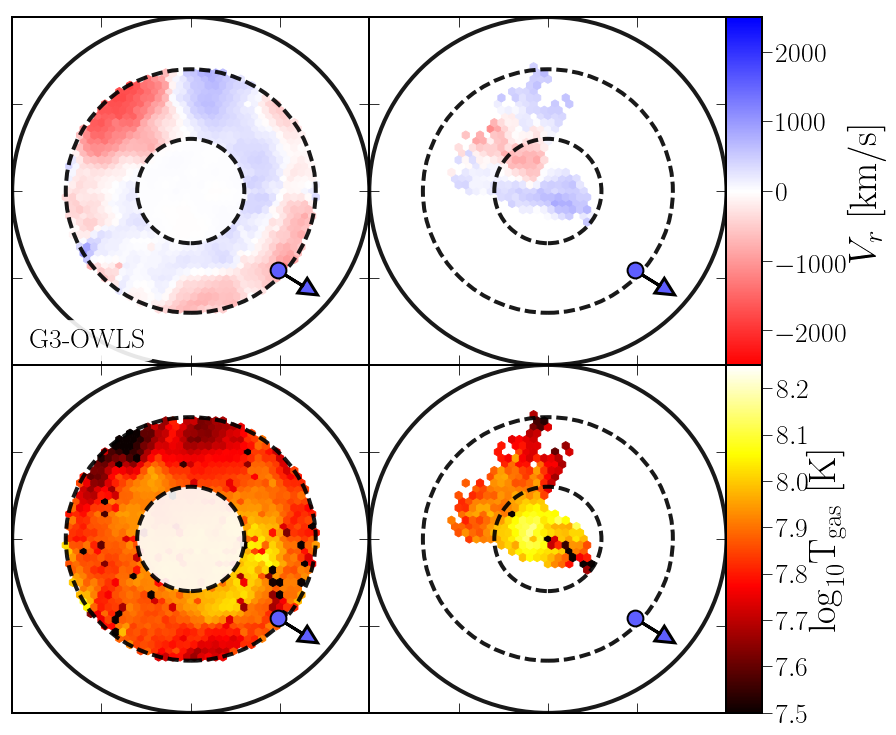}
  \includegraphics[width=0.49\textwidth, trim=0.0cm 0.0cm 0.0cm 0.0cm, clip=true]{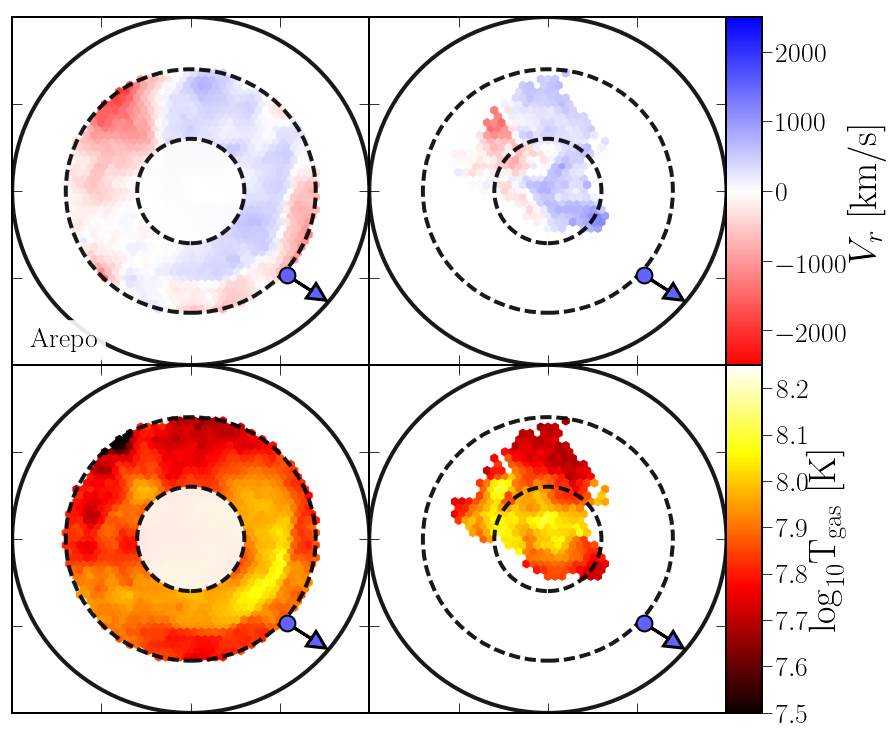}
  \caption{Gas properties around the inner shock in the {\small G3-OWLS} and {\small Arepo} full physics runs
    (upper and lower panels, respectively). We show, in projection, the mean radial velocity
    (upper panels) and temperature (lower panels) of gas in a shell around the shock feature (left panels)
    and the gas stripped from the outward going substructure (right panels) in the {\small Arepo} full physics run.
    We show also the position of the outward moving substructure and its direction of motion, with the point and
    arrow colour coded according to its radial velocity. The outer solid circle is at $R_{200}$, while the
    inner/outer dashed circles are at $0.4/0.8R_{200}$ and encompass the inner shock. For clarity, we do
    not include gas interior to the shell in projection.}
  \label{fig:outgoing_group_inner_shock_gas_properties}
\end{figure}

That a fast-moving, infalling, gas-rich substructure might generate a shock in the ICM is not surprising; but what
is the influence of the gas-poor substructure moving outwards? At $z$=0, it is moving with a radial velocity of
$\sim$1500~km/s; however, its position and velocity {\em do not} correspond to the inner
feature seen in the temperature and radial velocity profiles -- its radial velocity is too high, and it is at
too great a clustercentric radius. By tracking its orbit to earlier times, we can say that it had a mass on
infall of $\sim6\times10^{13}$~M$_\odot$ and it was gas-rich. At its pericentric passage of $\sim0.15\,R_{200}$,
it deposited a gas mass of $\sim8\times10^{12}$~M$_\odot$ in $\sim50$~Myrs as it moved at speeds of
$\sim2750$~km/s, and led to its baryon fraction plummeting from $\gtrsim0.5 \Omega_b/\Omega_m$ to
$\ll\Omega_b/\Omega_m$. It is this stripping event and the gas associated with it that produces the shock
at $\sim(0.5-0.6)R_{200}$. 

To illustrate these observations, in Figure \ref{fig:outgoing_group_inner_shock_gas_properties} we show the
projected radial velocities and temperatures (upper and lower panels respectively) of gas in the vicinity of
the inner temperature enhancement (left panels), and gas stripped from the outward moving substructure (right panels).
To show the generality of our results and the consistency between code behaviour, we consider two representative cases -
{\small G3-OWLS} at the top, and
{\small Arepo} at the bottom. Inspection of the left panels makes clear an arc of shock heated gas, both trailing the
outwardly moving substructure
and colliding with infalling gas; the right panels reveal that the shocked cluster gas is propelled outwards by the
cold, dense, fast moving gas stripped from the outwardly moving cluster. The fastest outwardly moving portion of
this stripped gas is spatially coincident with the inner edge of the shock, pushing material in front of it, and it is
this outwardly moving wake that shock heats when it encounters infalling gas. 

\subsection{The role of the cosmic web}
\label{ssec:lss}

A consistent feature of the results in \S~\ref{ssec:spherical_averages} is the
presence of an enhancement in gas density and temperature at
$\sim(2-3)\,R_{200}$. In contrast to the transient enhancements with $R_{200}$,
this feature is long-lived and we have argued that it is likely associated with
gas accretion from the cosmic web. We have also demonstrated that the massive
substructures whose passage through the ICM drives shocks have been accreted onto the
cluster from a preferential direction, which we interpreted as a signature for
infall along filaments.

We use {\small DisPerSe} \citep[cf.][]{2011MNRAS.414..350S} to characterise
filamentary structure in the vicinity of the cluster. {\small DisPerSe}
computes a discrete density field from a Delaunay
tessellation of particles and identifies filaments as ridges connecting maxima through saddle points in the
underlying density field. The result is a ``skeleton'' of the density field, consisting of a fully connected
network of segments, which can be smoothed (by pairwise averaging of neighbouring segments) to focus on a given
spatial scale\footnote{The skeleton can be smoothed on a given spatial scale, or it can filtered on a
  {\it persistence} threshold. Persistence characterises the robustness of the topology (i.e.
  number of holes, tunnels, threads, etc.) of the density field above a given excursion threshold, which can be
  interpreted as a measure of the significance of topological features, similar to the signal-to-noise ratio
  in observations. Note that persistence only characterizes robustness of pairs of critical points (such as
  maximum-saddle point height). We can however also derive a continuous robustness ratio that directly relates
  to the contrast of filaments (arcs between the critical points) relative to their background
  \citep{2011MNRAS.414..350S}. This allows us to distinguish between strong sustained filaments connecting
  high contrast nodes (high persistence level, high robustness ratio) and fainter, possibly more transient,
  features (low persistence level, low robustness ratio).}. Here we compute the skeleton using both dark matter
and gas in cubes of size 4 and 20 $h^{-1}$ Mpc on a side, centered on the location of the maximum density of the
cluster.

\begin{figure*}
  \centerline{
    \includegraphics[width=1.8\columnwidth]{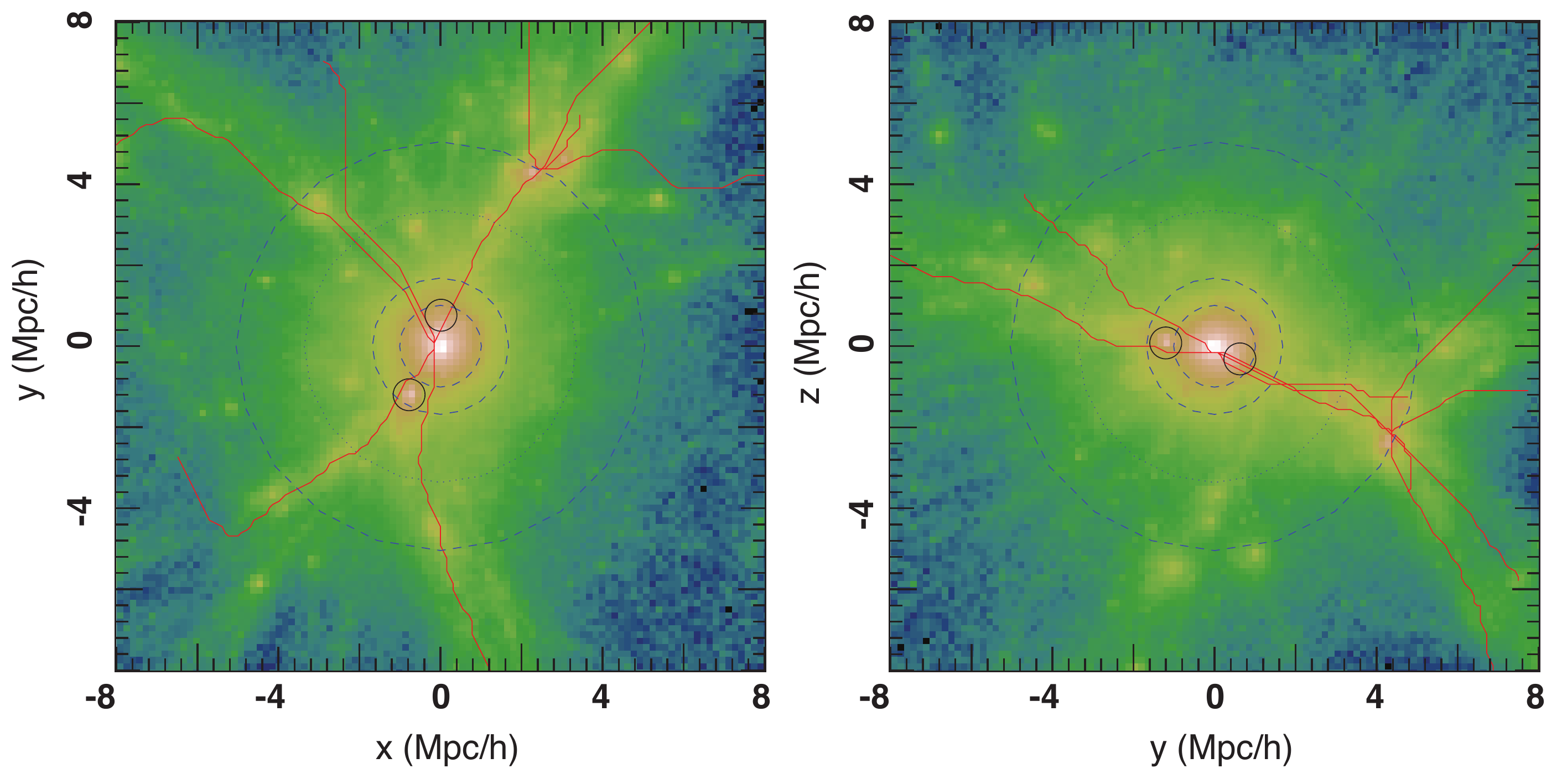}
  }
  \caption{Large-scale cosmic web connectivity in {\small G3-OWLS}. Projected gas density centred on
    the cluster
    within a $3 \, R_{\rm 200}$ thick slice in the $x$-$y$ (left panel) and $y$-$z$ (right panel) planes,
    colour-coded by the logarithm of the density. As described in the text, the solid red lines
    show the network of filaments recovered by {\small DisPerSe} \citep[cf.][]{2011MNRAS.414..350S}
    while the dashed circles indicate the locations of features evident in the spherically averaged
    radial velocity profiles, and solid circles correspond to the two massive substructures investigated in
    the previous subsection.}
  \label{fig:cosmic-web}
\end{figure*}

Figure~\ref{fig:cosmic-web} gives a visual impression of the cosmic web in which the cluster is embedded.
Solid red lines show a $3 h^{-1} {\rm Mpc}$ wide slice of the network of filaments with a persistence threshold
$c$=0.01 in a $20\,h^{-1}{\rm Mpc} $ cube smoothed onto a $128^{3}$ mesh, consecutively trimmed of arcs with
robustness smaller than the mean plus 1 standard deviation and smoothed over a $3 h^{-1} {\rm Mpc}$ spatial scale.
While these results are calculated for the gas, we note that we recover a very similar skeleton on large scales
using the dark matter density. Dashed circles indicate the locations of the features in the radial velocity profiles,
while the solid circles correspond to the gas-poor outwardly moving substructure (lower in $x$-$y$, left in $y$-$z$) and
gas-rich infalling substructure respectively.

The result is visually striking - Figure~\ref{fig:cosmic-web} makes clear how
the trajectories of both substructures can be traced to larger-scale filaments,
while also highlighting the connection between the enhancements in gas densty
and temperature, as well as the spherically averaged negative radial velocity,
at $\sim(2-3)\,R_{\rm 200}$, to the merging of several smaller filaments along
the direction of the most robust one. This explains why the location of this
feature should be long-lived, albeit varying in magnitude,
whereas the shocks associated with infalling substructures within $R_{200}$ should
be short-lived and dynamic.

\medskip

\section{Summary}
\label{sec:summary}

We have used a suite of cosmological hydrodynamical zoom simulations of a
single galaxy cluster, run with a range of astrophysical codes and galaxy
formation models as part of the {\small nIFTy} comparison project
\citep[see][]{2016MNRAS.457.4063S,2016MNRAS.tmp..582S}, to study how
the thermodynamical properties of the intra-cluster medium (ICM) in a
galaxy cluster's outskirts ($0.3 \leq R/R_{200} \lesssim 3$) at $z$=0
are influenced by the accretion of substructure from the cosmic web. Using
the radial profiles of gas density, temperature, and radial velocity,
we highlighted the presence of enhancements in gas density and temperature,
which we noted are associated with shocks (as measured by the Mach number)
and which are commonly interepreted as signatures of the passage of substructure
\citep[see, e.g.,][]{2013MNRAS.428.3274Z,2017ApJ...849...54L}. We showed
how such features within $R_{200}$ can be linked
to substructure evident in projected gas density and temperature maps,
and can be connected to the passage of substructures through the ICM.

In particular, we looked in detail at two massive substructures
- one gas-rich substructure infalling on a preferentially radial orbit with
a relative speed of $-1700$km/s, producing the clear signature of a shock
in the direction of its motion; the other gas-poor substructure, moving outwards
with a velocity of $1500$km/s, and gravitationally influencing gas in its
vicinity. By
constructing the cosmic web in the vicinity of the cluster, we established
that the cluster accretes from two dominant filaments, but that a network of
filaments combine at $\sim(2-3)\,R_{200}$ to give rise to density and
temperature enhancements, associated with the infall of gas and dark matter
along the cosmic web. Because the trajectories followed by the
substructures that give rise to the shocks within $R_{200}$ are fixed by the
filaments from which they were accreted, this suggests a way to reconstruct
the infall time and direction of massive substructures from observational data 
\citep[e.g. radio sychrotron measurements of termination shocks associated with the infalling galaxy substructures][]{2011MNRAS.412....2B}.

We note general qualitative agreement between the selection of astrophysical
codes and galaxy formation models (non-radiative and full physics) used in
this paper in terms of their recovery of spherically averaged cluster gas profiles,
at the level of $\sim 25\%$ or 0.1 dex between runs. Over the
radial range $0.3\,R_{200} \leq R \lesssim R_{200}$, we find broad agreement
between predictions for the locations of enhancements in gas density and
temperature, and when and where shocks occur, but 
differences between predictions for the sizes of enhancements and how strong
shocks - as gauged by the root-mean-square Mach number, $M_{\rm rms}$ - should be, most notably
between full physics runs. This is as we would
expect. Both gas and substructure accretion onto the cluster will be driven
by the large-scale gravitational field, which dictates when and where, and
we expect excellent agreement between the codes because of our requirement
that they are aligned \citep[cf.][]{2016MNRAS.457.4063S}, whereas shock
strength will
be driven by a number of factors, including resolution and choice of hydro
solver, and so we expect greater variation between codes. 

Although we have considered only a single galaxy cluster, albeit multiple
realisations with different codes and galaxy formation models, our results
imply that the dynamical effects of accreting substructure on ICM properties
will need to be accounted for when making accurate predictions for next
generation
surveys, such as e.g. the {\small Square Kilometre Array} in the radio continuum,
tracing synchrotron emission
\citep[e.g.,][]{2015aska.confE.104G,2015aska.confE..97V}, and
{\small eRosita} \citep[e.g.,][]{2012arXiv1209.3114M},
{\small X-Ray Imaging and Spectroscopy Mission (XRISM)}
\citep[e.g.,][]{2018SPIE10699E..22T},
and {\small ATHENA} \citep[e.g.,][]{2013arXiv1306.2307N} in X-ray.
Conversely, these surveys can open up a new view of the merger and accretion
histories of clusters.

In summary, we have investigated in detail how the passage of massive
substructures influence the ICM at large cluster-centric radius and shown
how it connects to features in radial ICM profiles, which previously have been
quantified in a statistical sense
\citep[e.g.,][]{2013MNRAS.428.3274Z,2014ApJ...791...96R,2014ApJ...782..107N,2015ApJ...806...68L,2018PASJ...70...51O,2018MNRAS.476...56Z}. Our case study
of the two massive substructures shows that the manner in which such systems drive shocks into
the ICM is not uniform; in one case, we have the classical picture
of a gas-rich, infalling, substructure driving a shock, while in the other, it is stripped
gas, which was once associated a now gas-poor, outgoing, substructure, is driving
a shock. Finally, we have verified that the predictions of different astrophysical codes and galaxy formation models are in broad agreement, predicting
spherically averaged radial properties of the ICM that agree at the 25\% (0.1 dex) level. 

\section*{Acknowledgments}

\noindent
The authors thank the anonymous referee for their comments.
The authors thank Alexander M. Beck (AMB) and Giuseppe Murate (GM) for their
contributions to the nIFTy simulations dataset.
CP acknowledges the support of an Australian Research Council
(ARC) Future Fellowship FT130100041. CP and AK acknowledge the support of
ARC Discovery Project DP140100198. CP, PJE, WC, and AK acknowledge the
support of ARC DP130100117. CP acknowledges the support of the ARC Centre
of Excellence in All-Sky Astrophysics (CAASTRO) through project number
CE110001020, and the ARC Centre of Excellence in All-Sky Astrophysics in
3 Dimensions (ASTRO 3D) through project number CE170100013. CW acknowledges
the support of the Jim Buckee Fellowship in Astrophysics at ICRAR/UWA.
AK is supported by the {\it Ministerio de Econom\'ia y Competitividad} and the
{\it Fondo Europeo de Desarrollo Regional} (MINECO/FEDER, UE) in Spain through
grant AYA2015-63810-P and the Spanish Red Consolider MultiDark FPA2017-90566-REDC.
He further thanks Pavement for Brighten the Corners.
GM acknowledges support from the PRIN-MIUR 2012 Grant "The Evolution of
Cosmic Baryons" funded by the Italian Minister of University and Research,
and from the PRIN-INAF 2012 Grant "Multi-scale Simulations of Cosmic
Structures", funded by the Consorzio per la Fisica di Trieste. EP acknowledges
support from the Kavli foundation and European Research Council (ERC) grant,
"The Emergence of Structure during the epoch of Reionization".
\smallskip

\noindent The {\small MUSIC} simulations were performed on the Marenostrum
Supercomputer at BSC using a computing time allocation granted by the Red
Espa\~{n}ola de Supercomputacion. The {\small Arepo} simulations were
performed with resources awarded through STFC's DiRAC initiative. The Arepo
simulation was performed on the DiRAC (\texttt{www.dirac.ac.uk}) systems:
Data Analytic at the University of Cambridge [funded by BIS National
  E-infrastructure capital  grant (ST/K001590/1), STFC capital  grants
  ST/H008861/1  and  ST/H00887X/1, and   STFC   DiRAC Operations   grant
  ST/K00333X/1]. DiRAC is part of the National E-Infrastructure.
\smallskip

\noindent The authors thank the Instituto de Fisica Teorica (IFT-UAM/CSIC)
in Madrid, via the Centro de Excelencia Severo Ochoa Program under Grant
No. SEV-2012-0249, whose support of the ``nIFTy Cosmology: Numerical
Simulations for Large Surveys'' workshop in mid-2014 facilitated the
initiation of this work. 
\smallskip

\noindent The authors contributed to this paper in the following
ways: CP, PJE, \& CW formed the core team that organized and analyzed
the data, made the plots and wrote the paper. AK, FRP, GY \& CP
organized the nIFTy workshop at which this program was initiated.
GY supplied the initial conditions. STK, AMB, IGMcC, GY, GM, \& EP
performed the simulations using their codes. All authors read and
commented on the paper.

\smallskip
\noindent This research has made use of NASA's Astrophysics Data
System (ADS), the arXiv preprint server, and the {\small py-sphviewer}
package of \citet{py-sphviewer}.

\vspace{1cm} \bsp

\bibliographystyle{mn2e}

\appendix

\section{Simulation codes} \label{A:codes}

\noindent \paragraph*{{\small Arepo}} (Puchwein)

\noindent The basic {\small Arepo} code \citep[cf.][]{2010MNRAS.401..791S}
utilises a TreePM gravity solver and a finite-volume Godunov
scheme on an unstructured moving Voronoi mesh to solve the equations of
hydrodynamics. Two galaxy formation prescriptions are used in the full
physics runs; the first ({\small Arepo-IL}) is used in the
\emph{Illustris Simulation}, includes AGN feeding and feedback, and is
described in \citet{2013MNRAS.436.3031V,2014MNRAS.444.1518V}.
The second ({\small Arepo-SH}) uses the standard \citet{2003MNRAS.339..289S}
scheme, without AGN, and matches that used in {\small G3-MUSIC}.

\smallskip

\noindent\paragraph*{{\small G2-X}} (Kay)

\noindent {\small G2-X} is built on the public version of
{\small GADGET2} \citep{2005MNRAS.364.1105S}, using the
standard cubic spline kernel with 50 neighbours. The adopted
galaxy formation prescription is descrived in detail in
\citet{2014MNRAS.445.1774P}, but it can summarised as follows;
radiative cooling follows the \citet{1992MNRAS.257...11T} prescription;
stars form following \citet{2008MNRAS.383.1210S} at a rate fixed by
the Schmidt-Kennicutt relation \citep{1998ApJ...498..541K},
and produce feedback using a prompt thermal Type II supernova model;
and the modelling of AGN feeding and feedback is based on
\citet{2009MNRAS.398...53B}.

\smallskip

\noindent\paragraph*{{\small G3-MUSIC}} (Yepes)

\noindent {\small G3-MUSIC} is built on {\small GADGET3}, which
derives from the public version of the {\small GADGET2} code of
\citet{2005MNRAS.364.1105S}, with improvements in time-stepping
and domain decomposition. It employs the entropy-conserving
formulation of SPH described in \citet{2002MNRAS.333..649S},
with a spline kernel \citep{1985A&A...149..135M} and artificial
viscosity as modelled in \citet{1997JCoPh.136..298M}. It uses two models -
the standard \citet{2003MNRAS.339..289S} scheme ({\small G3-MUSIC-SH})
and the \citet{2011MNRAS.410.2625P} scheme ({\small G3-MUSIC-PS}) -
neither of which include AGN, in the full physics runs.

\smallskip

\noindent\paragraph*{{\small G3-OWLS}} (McCarthy, Schaye)

\noindent {\small G3-OWLS} is built on {\small GADGET3} and
employs the standard entropy-conserving SPH scheme of
\citet{2002MNRAS.333..649S} with a cubic spline kernel
of 48 neighbours. The galaxy formation prescription used
in the full physics run is presented in extensive detail
in \citet{2008MNRAS.383.1210S}, \citet{2008MNRAS.387.1431D},
\citet{2009MNRAS.393...99W}, \citet{2009MNRAS.398...53B},
\citet{2009MNRAS.399..574W}, \citet{2010MNRAS.402.1536S},
and encompasses modelling of radiative cooling, star formation,
stellar evolution, stellar feedback, and AGN feeding.

\smallskip

\noindent \paragraph*{G3-X} (Murante, Borgani, Beck)

\noindent {\small G3-X} is built on {\small GADGET3} and developed by 
\citep{2016MNRAS.455.2110B} to include a Wendland $C^4$ kernel with
200 neighbours \citep[cf.][]{2012MNRAS.425.1068D}, artificial
conductivity to promote fluid mixing following \cite{2008JCoPh.22710040P}
and \cite{2013MNRAS.436.2810T}, but with an additional limiter for
gravitationally induced pressure gradients. The full physics run
adopts the cooling prescription of \citet{2009MNRAS.393...99W};
heating via a uniform UV background as in \citet{2001cghr.confE..64H};
star formation and chemical evolution as in \citet{2007MNRAS.382.1050T};
stellar feedback in the form of supernovae as in \citet{2003MNRAS.339..289S};
and AGN feedback as in \citet{2015MNRAS.448.1504S}.

\label{lastpage}

\end{document}